\documentstyle[eqsecnum,preprint,aps]{revtex}
\begin{document}
\draft

\title{PATTERN FORMATION IN LASER INDUCED MELTING}

\author{Chuck Yeung and Rashmi C.\ Desai}

\address{Department of Physics, University of Toronto, Toronto, Ontario
M5S 1A7 CANADA}

\date{\today}
\maketitle
\begin{abstract}
A laser focussed onto a semiconductor film can create
a disordered lamellae pattern of coexisting molten-solid regions.
We present a continuum model based on the higher reflectivity of the
molten regions.  For large latent heat, this model becomes equivalent
to a model of block copolymers. The characteristic wavenumber
of the lamellae is that marginally stable to slow variations in
the orientation (the zig-zag instability) and is obtained via
systematic expansions from two limits.  The lamellae can also be
unstable to the zig-zag instability and Eckhaus instability (slow
variations in the wavenumber) simultaneously.  This instability is a
signal of dynamic steady states.  We numerically study the behaviour
after a quench.  The lamellar size agrees with the analytic
results and experiments.  For shallow quenches, locally parallel stripes
slowly straighten in time.  For deep quenches, a disordered lamellae
forms.  We construct the director field and determine the
orientational correlation length.  Near onset the correlation is fixed
by the system size.  Far from onset the correlation length saturates at
a finite value.  We study the transition to the time-dependent asymptotic
states with decreasing latent heat.
\end{abstract}
\pacs{47.54+r,44.30+v,5.70 Ln, 78.66-w, 64.60 Cn, 61.41+e\\
{\em submitted to Phys.\ Rev. E}


\section{Introduction}

A laser focussed onto a semiconductor film such as silicon can induce
melting.  At low laser intensity the incident light simply raises
the sample temperature.  At large laser intensity a uniform molten
surface is created.  However, the laser intensity can be adjusted
so melting is initiated but molten and solid regions coexist
\cite{BOSCH,HAWKINS,NEMANICH,PRESTON,DWORSCHAKJOSA,DWORSCHAK}. In this
case, the illuminated area contains both molten and solid
regions of size much smaller than the laser beam diameter.

The  coexisting molten-solid regions can arise from two
different mechanisms.  One possibility discussed by Preston et al.\
\cite{PRESTON,PRESTON89} is, due to the coherence of the laser beam,
there can be interference between the incident and surface scattered
fields.  This leads to spatially periodic power deposition and solid/melt
patterns that depend on the wavelength and orientation of the incident
light \cite{NEMANICH,PRESTON}.

In this paper we will be interested in a second mechanism. It was
conjectured that the non-uniformity is due to the increased reflectivity
of the melt \cite{HAWKINS,JACKSON}. Following the argument of Hawkins
and Biegelsen \cite{HAWKINS}, assume that laser energy flux $J_{1}$ is
required to raise the solid sample to the melting temperature.  Due to
the higher reflectivity, the energy flux $J_{2}$ required to maintain
a homogeneous molten phase at that temperature will be much higher.
Therefore for energy fluxes between $J_{1}$ and $J_{2}$ a homogeneous
phase is not possible.  Instead  a heterogeneous surface is created
in which the molten regions are undercooled and the solid regions
superheated.  Although a free energy argument may not be justified for
the nonequilibrium steady state, Hawkins and Biegelsen \cite{HAWKINS}
balanced the surface energy due to the solid/liquid interfaces with
that due to the undercooling (superheating) of the melt (solid) phase.
They were then able to relate the pattern wavelength to the average
undercooling.  Jackson and Kurtze \cite{JACKSON}, on the other hand,
studied the stability of a periodic array of solid and molten stripes
using a phenomenological interface description.  They found that, if the
wavelength of the pattern is within a specific band, the pattern is stable
to perturbations in the orientation of the stripes.  This
thermally controlled regime has also been observed experimentally and is
characterised by a disordered lamellar structure \cite{HAWKINS,DWORSCHAK}.

In addition to static patterns, the transition between the parallel
stripes and disordered lamellar phases have been studied.
The transition was found to be
reversible but displays hysteresis characteristic
of a first order transition  \cite{DWORSCHAK}.  Dynamical structures
have also been observed \cite{BOSCH,HAWKINS,NEMANICH,PRESTON} in which
the patterns evolves to some, possibly chaotic,
time-dependent asymptotic state.

In Section II we discuss our continuum model of laser induced melting.
The dynamics are given by a set of coupled partial differential equations
for the order parameter and temperature field.  The effect of the
incoming energy of the laser, the heat flow to the substrate and the
order parameter dependent reflectivity are included.
In Section III we show that, in the limit of large latent heat, this
model is equivalent to a model of block copolymers
\cite{LEIBLER,OHTA}.  In particular the static solutions are always
that of block co-polymers. Using this analogy,
the characteristic length-scale of the patterns is the one that minimises
the free energy. We obtain this length-scale  via
systematic expansions close to (weak segregation), and far away (strong
segregation) from onset (the parameters where pattern formation first
occurs).  These regimes are characterised by different dependence on the
control parameter \cite{OHTA,LIU,OONO} and by qualitatively different
patterns.  Near onset, locally parallel stripes are observed while,
far from onset, disordered lamellar structures are found.

Although we discuss a specific model, this analogy between seemingly
unrelated systems leads us to expect that our results
are relevant to many systems which favour phase separation
at short length-scales but has a mechanism suppressing macroscopic
phase separation. Examples include ferrofluids \cite{ROSENWEIG},
ferromagnets \cite{MOLHO}, monolayer films \cite{SEUL,ROLAND}, and
Raleigh-B\'{e}nard convection \cite{BENARD}.  In particular,
the patterns after a quench are characterised by two length-scales
corresponding to the length of the local domains (stripes
for the symmetric case) and the length on which the orientation of the
domains are correlated.  This additional length-scale has been suggested
for Raleigh-B\'{e}nard convection \cite{TONER,ELDER}.

Whereas, for large latent heat the dynamics are equivalent to that of
block copolymers, for other parameters the static state is never reached
and the system instead settles into a time-dependent asymptotic state.
In Section IV we discuss different stability criteria.  We consider
the linear analysis of the homogeneous state.  We derive the phase
diffusion equations \cite{CROSS,MANNEVILLE} describing the slowly
varying lamellae.  Using the phase diffusion equations we re-obtain the
equilibrium length-scale using the rigourous criteria that the selected
length-scale is the one marginally stable to the zig-zag instability
(i.e., marginally stable to slow variations in the orientation of the
stripes).  We therefore make concrete, and relate to one another, the
seemingly unrelated free energy \cite{HAWKINS} and interfacial stability
approaches \cite{JACKSON}.

In addition, we find a new instability is possible in which the lamellae
are unstable to both Eckhaus and zig-zag instabilities simultaneously.
This instability occurs at small latent heat and marks the limit of
metastability of the parallel lamellae.  We conjecture that this
instability is a signal of experimentally observed time-dependent
asymptotic states \cite{BOSCH,HAWKINS,DWORSCHAK}.

In Section V we present a numerical study.  First we consider static
steady states.  Previous simulations only probed the characteristic
lengthscale close to onset \cite{OONO,BAHIANA}.  We find agreement
with theory both near and far from onset.  To study the correlation of
the stripe orientation, we construct a director field pointing in the
direction of the stripes.  Near onset, the orientational correlation
length is much larger (possibly only limited by the system size) than
the pattern wavelength.  This leads to large regions of locally parallel
stripes.  Far from onset, the orientational correlation length is the
same order as the pattern wavelength.  A disordered lamellae is formed
in which the lamellae bend on the same length-scale as the lamellar
width.  The ratio of the correlation length divided by the pattern
wavelength decreases monotonicly with quench depth.  We speculate that
a nonequilibrium Kosterlitz-Thouless like transition exists separating
regimes with long range orientational order from those without.  We then
map out the phase boundary between static and time-dependent asymptotic
states.  The various regimes are described by the phase diffusion dynamics
and not by the linear dynamics around the unstable initial condition.

\section{Model Equations}

We introduce our model of laser induced melting.  Let $\phi$ be a
dimensionless order parameter field which characterises the solid or
liquid phases.	At coexistence we choose $\phi = 1$ for the liquid phase
and $\phi = -1$ for the solid phase.  The free energy is \begin{equation}
	F \{ \phi \} = \int d{\bf r} \; \left[ \rho_{0} \epsilon_{0}
	\left( f( \phi ) + \frac{ \xi_{0}^{2} }{2} \left|
	\mbox{\boldmath{$\nabla$}} \phi \right|^{2} \right) - \rho_{0}
	\frac{L}{2} u \phi \right].
\end{equation} Since there are two coexisting phases, $f(\phi)$
is required to be a dimensionless function of double well form with
minima at $\phi = \pm 1$ corresponding to the liquid and solid phases.
For simplicity we assume $f(\phi)$ to be symmetric.  The $|
\mbox{\boldmath{$\nabla$}} \phi |^{2}$ term penalises gradients in
$\phi$ and is required since, in the absence of driving, the homogeneous
state has the lowest free energy.  The last term is the coupling to the
reduced temperature $u = (T-T_{m})/T_{m}$ where $T_{m}$ is the melting
temperature.  For $u > 0$, the liquid phase is favoured while for $u <
0$ the solid phase has the lower free energy.  The physical constants
are (for simplicity we assume the two phases are symmetric):
$\rho_{0}$ is the mass per unit volume, $\epsilon_{0}$ is a constant with
units of energy/mass, $\xi_{0}$ is a microscopic interfacial width and
$L$ is the latent heat per unit  mass.	The surface tension is $\sigma
= \rho_{0} \epsilon_{0} \xi_{0} \bar{\sigma}$ where the dimensionless
surface tension, $\bar{\sigma}$, depends on the precise form of $f(\phi)$
and is of order unity.	For $f( \phi ) = -\phi^{2}/2 + \phi^{4}/4$,
$\bar{\sigma} = 2 \sqrt{2}/3$.

To describe the dynamics of $\phi$ we assume that $\phi$ obeys the
time-dependent Ginzburg-Landau equation \begin{eqnarray}
	\partial_{t} \phi( {\bf r}, t) & = & -\frac{1}{\rho_{0}
	\epsilon_{0} \tau_{0}} \frac{ \delta F }{ \delta \phi( {\bf r},
	t) }
		\label{eq:TDGL} \\
	& = & \frac{1}{ \tau_{0} }\left[ -\mu_{B}( \phi( {\bf r}, t) )
	+ \xi_{0}^{2} \nabla^{2} \phi( {\bf r}, t)
		+ \frac{L}{2 \epsilon_{0}} u( {\bf r}, t) \right],
		\nonumber
\end{eqnarray} where $\mu_{B} = \partial f/\partial \phi$ and  $\tau_{0}$
is a microscopic time-scale.

We also need the dynamics of the reduced temperature field $u$.
In actual experiments the geometry is not translationally invariant
\cite{BOSCH,HAWKINS,NEMANICH,PRESTON,DWORSCHAKJOSA,DWORSCHAK}. Here we
consider a simpler geometry in which the film sits on a substrate which,
in turn, is attached to a heat bath.  The thickness of the film is assumed
sufficiently small so that all variation in perpendicular direction can
be neglected.  (This will be true if the thermal conductivity of the
substrate is much smaller than that of the film.) The reduced temperature
field obeys \begin{equation}
	\rho_{0} C_{p} \, \partial_{t} u = K_{T} \nabla^{2} u -
	\frac{\rho_{0} L}{2 T_{m}} \, \partial_{t}  \phi + \frac{1}{h
	T_{m}} J_{in}(\phi) - \frac{1}{h T_{m}} J_{out}(u).
		\label{eq:TEMPDYNAMICS}
\end{equation} Here $K_{T}$ is the thermal conductivity, $C_{p}$ the
specific heat per unit mass and $h$ the thickness of the film.

The incoming flux is $J_{in}(\phi)  =  J_{0} ( 1 - R( \phi ) )$,
where $J_{0}$ is the energy flux of the laser and $R( \phi )$ is the
reflectivity which is higher in the molten phase. For simplicity we assume
that $R(\phi)$ is linear in $\phi$.  We write $J_{in}$ as \begin{equation}
	J_{in}(\phi)= J_{in}( \phi=0 ) - J_{0} \frac{\Delta R}{2} \phi,
\end{equation} where $\Delta R$ is the difference in the reflectivity
between the solid and melt phases.

The outgoing flux is $J_{out}(T) = (K_{s}/d) \left( T - T_{0} \right)$,
where $K_{s}$ is the thermal conductivity of the substrate, $d$ is some
length of the order of the substrate thickness and $T_{0} < T_{m}$
is the temperature at the heat sink.  We assume that the system is
close to the melting temperature so the dependence of $J_{out}$
on the local temperature is negligible .  The net flux
is then \begin{equation}
	J_{Total} = J_{in}(\phi)
	- J_{out}(T) = \Delta J - J_{0} \, \frac{\Delta R}{2} \phi.
\end{equation}
Here $\Delta J \equiv J_{in}(\phi=0) - J_{out}(T=T_{m})$
determines the fractions of the melt and solid phases.
There will be a 50-50 mixture  if $\Delta J =
0$.  Increasing $\Delta J$ will increase the fraction of the molten phase.

We introduce the dimensionless variables $t/ \tau_{0} \rightarrow t$,
$r/ \xi_{0} \rightarrow r$ and $u \, L/(2 \epsilon_{0}) \rightarrow u$.
The dynamical equations becomes \begin{eqnarray}
	\partial_{t}  \phi & = & -\mu_{B} + \nabla^{2} \phi + u,
	\label{eq:PHIDYNAMICS}
		\\
	\partial_{t} u & = & D \nabla^{2} u - \ell \, \partial_{t}
	\phi + \Delta j - r_{0} \; \phi, \label{eq:UDYNAMICS}
\end{eqnarray} with the dimensionless parameters, \begin{eqnarray}
	D = \frac{ \tau_{0} K_{T}}{\xi_{0}^{2} \rho_{0} C_{p}  }, & \;
	\; & \ell = \frac{L^{2}}{4 \epsilon_{0} T_{m} C_{p} },
				\nonumber \\
	\Delta j = \frac{ \tau_{0} L \Delta J}{2 \epsilon_{0} h T_{m}
	\rho_{0} C_{p}}, & & r_{0} = \frac{  \tau_{0} L J_{0} \Delta R
	}{4 \epsilon_{0} h T_{m} \rho_{0} C_{p}}.
\end{eqnarray}
We will use these rescaled equations as our working
model. There are four independent parameters.  We discuss primarily the
symmetric case, i.e., $\Delta j = 0$. Equations (\ref{eq:PHIDYNAMICS})
and (\ref{eq:UDYNAMICS}) with $r_{0} = 0, \Delta j = 0$ have been used
as models of phase ordering dynamics with conservation of order parameter
\cite{COLLINS}.

\section{Analogy with Block Co-Polymers}

\subsection{Dynamical Equation and Static States}

The fundamental feature of laser induced melting is that there is local
phase segregation but macroscopic phase separation is suppressed due
to an effective long range interaction. This is a feature common of
many physical systems.	Therefore we ask whether there is a relationship
between different models with this feature.  Here we show that
in the limit of large latent heat the dynamics are equivalent to that of a
model of block copolymers.  In particular, the static configurations are
always static configurations of the block copolymer model.  Using this
analogy we obtain the characteristic periodicity of the patterns using
systematic expansions around two limits.

A block copolymer is a linear polymer chain consisting of two subchains
$A$ and $B$ covalently bonded to each other.  As the repulsion between
$A$ and $B$ is increased there can be segregation of $A$ and $B$
chains.  However, due to the covalent bond macroscopic phase separation
is impossible.	Instead the static state consists of alternating $A$
and $B$ domains of mesoscopic size.

The scaled free energy for block copolymers  is
\cite{LEIBLER,OHTA} \begin{eqnarray}
	F\{ \phi \} & = & \int d {\bf r} \; \left[ f( \phi( {\bf r} )
	) + \frac{ 1 }{ 2 } | \mbox{\boldmath{$\nabla$}} \phi( {\bf r}
	) |^{2} \right]
	- \, \frac{B}{2 } \int d {\bf r} \, d {\bf r}' \; \phi( {\bf r}'
	) G( {\bf r}, {\bf r}' ) \phi( {\bf r} ),
\end{eqnarray}
where $B \sim 1/N^{2}$ with $N$ being the degree of
polymerization, $\phi$ is the difference in the local volume fraction
of $A$ and $B$ monomers, $G( {\bf r}, {\bf r}' ) $ obeys $\nabla^{2} G(
{\bf r}, {\bf r}' ) = \delta( {\bf r}- {\bf r}')$, and $f$ is a double
well coarse-grained free energy.   We can choose $f$ to be the same as that
of our laser induced melting model.  Assuming relaxational dynamics,
the dynamics for block copolymers is \cite{OONO,BAHIANA}
\begin{equation}
	\frac{1}{ \Gamma } \partial_{t}  \phi		= \nabla^{2}
	\frac{ \delta F }{ \delta \phi }
		=
	\nabla^{2} \left( \mu_{B} - \nabla^{2} \phi \right) - B ( \phi(
	{\bf r} ) - \bar{ \phi } ), \label{eq:DYNAMICCOPOLYMER}
\end{equation}
where $\mu_{B} = \partial f/\partial \phi$, $\bar{\phi}$
is the average value of $\phi$ and $\Gamma$ is a kinetic coefficient.
The static solution obeys
\begin{equation}
		0 =
	\nabla^{2} \left( \mu_{B} - \nabla^{2} \phi \right) - B ( \phi(
	{\bf r} ) - \bar{ \phi } ).  \label{eq:STATICCOPOLYMER}
\end{equation}
There are two regimes.  For $B$ very close to
the onset of microphase separation, the pattern lengthscale grows as
$B^{-1/4}$ \cite{LEIBLER,OHTA}.  This is called the weak segregation
regime.  Further from onset is the strong segregation regime in which
the domain grow as $B^{-1/3}$ \cite{OHTA,LIU,HASHIMOTO}. In this
regime one observes a disordered lamellar structure unless the
overall direction is fixed by flow or by boundary
conditions \cite{BAHIANA} or if the system is gradually annealed
\cite{CHAKRABARTI}.

Now we reconsider the laser induced melting model.  We can use Eq.\
(\ref{eq:PHIDYNAMICS}) to solve for $u$ \begin{equation}
	u = \mu_{B}(\phi) - \nabla^{2} \phi+ \partial_{t} \phi.
\end{equation} Substituting  into Eq.\ (\ref{eq:UDYNAMICS}) gives
\begin{eqnarray}
	\partial^{2}_{t} \phi + \ell \partial_{t} \phi + \frac{ d \mu_{B}
	}{d \phi} \partial_{t} \phi - (D+1) \nabla^{2} \partial_{t}
	\phi & \nonumber \\ = D \nabla^{2} \left( \mu_{B} - \nabla^{2}
	\phi \right) + \Delta j - r_{0} \phi. &
\end{eqnarray} For $\ell \gg 1, \ell \gg D$, the LHS is dominated by
the $\ell \partial_{t} \phi$ term and the dynamical equation becomes
\begin{equation}
	\frac{\ell}{D} \partial_{t} \phi
		=
	\nabla^{2} \left( \mu_{B} - \nabla^{2} \phi \right) -
	\frac{r_{0}}{D} \left( \phi - \frac{\Delta j}{r_{0}} \right),
\end{equation} and is therefore equivalent to the {\em dynamics} of
block copolymers (Eq.\ (\ref{eq:DYNAMICCOPOLYMER})) with $\Gamma =
D/\ell$, $B = r_{0}/D$ and $\bar{\phi} B = \Delta j/D$.  In addition,
independent of the values of $\ell$ and $D$, the static solutions are
extremums of the block copolymer free energy.

Figure \ref{fig:PICTURE} shows the $\phi$ configurations obtained from
a simulation of the laser induced melting model.  We will discuss
the simulations in detail in later sections.
Two regimes exists with qualitatively different patterns.  In analogy with
block copolymers we call these the strong and weak segregation regimes.
On the left, the parameters corresponds to the weak segregation regime
near onset.  There are long parallel stripes straightening in time.
On the right, the parameters are further from onset.  This corresponds
to the strong segregation regime in block copolymers.  Here a disordered
lamellae structure is found.  That is, rather than straight stripes,
there is a more complicated interconnected structure.  In this regime a
cross-section of the order parameter profile would show domains of the
two phases separated by a thin interfacial region over which the value
of $\phi$ changes from +1 to $-$1.

Using this analogy we will use known results for block copolymers to aid
in our study of `laser induced melting'.  We will also use the `laser
induced melting' model to shed some light on block copolymers.	However,
the dynamics for laser induced melting is richer.  For example,
for small $\ell/D$, the static solution
is not obtained.  Instead a time-dependent asymptotic state is reached
which can be chaotic.

\subsection{Energy Minimisation}

The equilibrium length-scale is the wavelength of the one-dimensional
solution which minimises the free energy. We assume a one-dimensional
solution of wavenumber $k = 2 \pi/\lambda$.  The free energy
density is \cite{LIU}
\begin{eqnarray}
	& & \frac{1}{\lambda} \int^{\lambda/2}_{-\lambda/2} dx \;
		\left[ f( \phi ) + \frac{ 1 }{ 2 } | \partial_{x} \phi
		|^{2} \right]
	- \; \frac{ B}{2 \lambda} \int^{\lambda/4}_{-\lambda/4} dx \;
	\int^{\lambda/4}_{-\lambda/4} dx' \;
	 | x - x' | \phi(x) \phi(x'),
\end{eqnarray}
where we choose $\phi(x) = -\phi(-x)$.  In this subsection
we obtain the equilibrium length-scales via
systematic expansions from the strong and weak segregation limits.
We also show that in the strong segregation limit, this is equivalent
to a modified form of the free energy arguments of Hawkins and Biegelsen
\cite{HAWKINS}.

\subsubsection{Strong segregation limit}

For small $B$ the equilibrium length-scale was obtained by Ohta and
Kawasaki using a variational method by assuming a form for $\phi(x)$.
We will show that this result can be justified via a systematic expansion
in the small parameter $1/\lambda$.  As shown in Appendix A, $\phi(x)
= \pm 1 + {\cal O}( \lambda^{-1})$ in the two bulk phases. Near the
interface $\phi(x) = \phi_{0}(x) + {\cal O}(\lambda^{-2})$ with $\phi_{0}$
being the planar interfacial profile for $B = 0$. We can now calculate
the free energy density.  There are two interfaces per period so the
local term in the free energy density is $2 \bar{\sigma}/\lambda +
{\cal O}(\lambda^{-2})$.  The long range term is $B \lambda^{2}/ 96 +
{\cal O}( \lambda^{-3} )$.  The free energy density is \begin{equation}
	\frac{F\{ \phi \}}{\lambda} =	\frac{ 2 \bar{\sigma} }{ \lambda
	}	       + \frac{\lambda^{2} B }{ 96 } + {\cal O}\left(
	\frac{1}{\lambda^{2}} \right).
\end{equation} (Here $B$ is ${\cal O}(\lambda^{-3})$.)	Minimising $F$
w.r.t.\ to $\lambda$ gives the wavelength of the one-dimensional solution
with the lowest free energy as \begin{equation}
	\lambda^{*} = \left( \frac{96 \bar{\sigma}}{B}	\right)^{1/3}
		+ {\cal O}( B^{0} ).
	\label{eq:LAMBDASTRONG1}
\end{equation} This is in agreement with the previous strong segregation
result for copolymers \cite{OHTA,LIU,OONO}.  Higher order corrections
can also be obtained to give \begin{equation}
	\lambda^{*} = \left( \frac{96 \bar{\sigma}}{B}	\right)^{1/3}
		+ \frac{8 \chi \bar{\sigma}}{5} + {\cal O}( B ).
	\label{eq:LAMBDASTRONG2}
\end{equation} where $\chi^{-1} = d \mu_{B}/ d \phi |_{\phi=1}$.
The wavelength for the laser induced melting model is obtained by
replacing $B$ by $r_{0}/D$.

We now reconsider the free energy argument of Hawkins and Biegelsen
\cite{HAWKINS}.  In the strong segregation limit, the free energy
density due to interfaces is $2 \bar{\sigma}/\lambda$.	They balanced
this with the change in the entropic contribution $T S$ (where $S$
is the entropy) due to the superheating and undercooling of the bulk.
In terms of the rescaled variables this contribution
is $-\frac{1}{\lambda} \int^{\lambda}_{0} dx \, u(x) \phi(x) =
\langle \, | u | \, \rangle$.  Balancing these two factors they obtained
$\lambda \langle \, | u | \, \rangle \approx 2 \bar{\sigma}$.  Note,
however, the static solution gives $u = (r_{0}/D) \nabla^{-2} \phi$,
so that the long range term in block co-polymer free energy is exactly
this bulk entropic contribution.  Therefore we have justified Hawkins
and Beigelsen's approximate analysis.  In fact, we can obtain the result
(Eq.\ \ref{eq:LAMBDASTRONG1}) if we extend their argument by explicitly
calculating $\langle |u| \rangle$ and minimising the resulting free
energy density.

For direct comparison with experiment the rescaling we have chosen
is somewhat confusing in the small $r_{0}$ limit.  In terms of
the dimensional unscaled variables we find (leading order only)
\begin{equation}
	\lambda^{*} = 4 \left( \frac{ 6 \sigma	h T_{m} K_{T} }{
		\rho_{0} L J_{0} \Delta R  } \right)^{1/3}.
\end{equation} So that the selected wavelength is independent of
the microscopic parameters $\xi_{0}$, $\epsilon_{0}$ and $\tau_{0}$.
As we will discuss in the next section this is $(6/27)^{1/3}$ of the
result obtained by Jackson and Kurtze using a stability analysis of a
periodic array of interfaces.  Our result is in very close agreement with
the experimental result of Dworschak and van Driel \cite{DWORSCHAK},
although, due to the difference in the experimental geometry and the
lack of symmetry between solid and liquid phases in the experiment,
a very detailed comparison is not appropriate.

\subsubsection{Weak segregation limit}

We can also find the wavelength minimising the free energy in the weak
segregation regime.  The homogeneous state first becomes unstable at $B =
B_{0} = 1/4$ and wavenumber $k = k_{0} = 1/\sqrt{2}$.  Near onset, we can
expand the static solution in orders of $\epsilon$ where $\epsilon^2 =
B_{0} - B$, \begin{equation}
	\phi(x) = \epsilon A_{k} \sin(k x) + \epsilon^{3} A_{3k} \sin(
	3 k x ) + {\cal O}(\epsilon^{5}),
		\label{eq:AMPSTATIC}
\end{equation} with $A_{k}^{2} = (8/3) \left( 1 - 2 \left( (k -
k_{0})/\epsilon \right)^{2} \right)$ and $A_{3k} = 9 A_{k}^{3}/128$.
Here $k - k_{0}$ is assume to be ${\cal O}(\epsilon)$.	In this limit
the free energy density is \begin{eqnarray}
	\frac{F\{\phi\}}{\lambda} & = & - \frac{\epsilon^{2} A_{k}^{2}}{4}
		\left( 1 - k^{2} \right) + \frac{ \epsilon^{4} 3 A_{k}^{4}
		}{ 32 } \left( 1 - 9 k^{2} \right)
			\nonumber \\
	& & \; + \, \frac{\epsilon^{2} B A_{k}^{2} }{ 4 k^{2} }
		+ {\cal O}(\epsilon^{6}).
\end{eqnarray} The first two terms correspond to the short range parts
of the free energy density while the last term is the long range portion.
We minimise this expression w.r.t.\ $k$ to give \begin{equation}
	k^{*4} = B + {\cal O}(\epsilon^{4}),
\end{equation} which is in agreement with the previous weak segregation
result \cite{LIU,OONO,BAHIANA}.  The next order correction can also be
calculated (one needs to keep terms of order $\epsilon^6$ in the free
energy) to give \begin{equation}
	k^{*4} = B \left( 1- \frac{5}{16} \epsilon^{4} \right) + {\cal
	O}(\epsilon^{6}).
\end{equation} We see that the fourth order correction is fairly small.
In the weak segregation limit the selected wavelength depends on the
microscopic parameters. In terms of the dimensional unscaled variables,
the leading order result is \begin{equation}
	k^{4} = \frac{ L J_{0} \Delta R }{ 4 \epsilon_{0} \xi_{0}^{2}
			h T_{m} K_{T} }.
\end{equation} This limit cannot be obtained from the previous interface
description.

\section{Stability Analysis}

\subsection{Linear Stability Analysis}

Since a Liapunov functional exists only in the large $\ell$ limit,
the laser induced melting model has much richer dynamics
than the block copolymer model.  Some insight can be
obtained through a variety of stability analysis.

The homogeneous solution of Eq.\ (\ref{eq:PHIDYNAMICS}) and
(\ref{eq:UDYNAMICS}) is $\phi^{*} = \Delta j/ r_{0}$, $u^{*} = \left.
\mu_{B} \right|_{\phi=\phi^{*}}$.  We assume infinitesimal perturbations
$\delta \phi_{k} = \delta_{\phi}  \exp(\omega_{k} t + i {\bf k}\cdot {\bf
r} )$ and $\delta u_{k} = \delta_{u}  \exp( \omega_{k} t + i {\bf k}\cdot
{\bf r} )$. Solving the dynamical equations (\ref{eq:PHIDYNAMICS}) and
(\ref{eq:UDYNAMICS}) to first order in $\delta$ gives, \begin{equation}
	\omega_{k}  = -\frac{\gamma_{k}'}{2} + \frac{ |\gamma_{k}'|}{2}
	\left( 1 - \frac{ 4( r_{0} + \gamma_{k} D  k^{2} ) }{
	\gamma_{k}'^{2} } \right)^{1/2} \label{eq:LINEAR},
\end{equation} where $\gamma_{k} = \mu_{B}' + k^{2}$ with $\mu_{B}' = d
\mu_{B} / d \phi |_{\phi=\phi^{*}}$ and $\gamma_{k}' = \ell + \gamma_{k}
+ D k^{2}$.

For $\mu_{B}' > 0$, the real part of $\omega_{k}$ is always negative and
the homogeneous state is linearly stable.  In this case, the initial state
is outside the spinodal.  For $\mu_{B}' < 0$ the linear dynamics can be
divided into several classes.  For $\ell < -\mu_{B}'$, the homogeneous
system is linearly unstable to small wavenumber perturbations ($k
\rightarrow 0$).  If $4 r_{0} > ( \ell + \mu_{B}')^{2}$ the instability
will be oscillatory.

If $\ell > -\mu_{B}$ the system is linearly stable to small wavenumber
perturbations.	If, in addition, $r_{0} > \mbox{max} \; (-\gamma_{k} D
k^{2} ) = D (\mu_{B}'/2)^{2}$ the homogeneous state is linearly stable to
perturbations of all wavenumbers.  For $r_{0} < D ( \mu_{B}'/2)^{2}$ the
system will be unstable in a finite band of wavenumbers, \begin{equation}
	 -\mu_{B}' -\sqrt{ (\mu_{B}')^2 - \frac{4 r_{0}}{D}}
	< 2 k^{2} <
	 -\mu_{B}' +\sqrt{ (\mu_{B}')^2 - \frac{4 r_{0}}{D}} .
\end{equation} For $\Delta j = 0$ and $\mu_{B} = - \phi + \phi^{3}$,
the conditions are (i) unstable at $k = 0$  if $\ell < 1$, (ii) stable
to all $k$ if $r_{0}/D > 1/4$ and (iii) if $r_{0}/D < 1/4$ then unstable
in a band of wavenumber $k$ with $1/2 - \sqrt{ 1/4 - r_{0}/D } < k^{2}
< 1/2 + \sqrt{ 1/4 - r_{0}/D }$.

\subsection{Phase Diffusion Description}

\subsubsection{Stability of lamellae}

The state minimising the block copolymer free energy corresponds to
parallel stripes.  The approach to this state is via locally parallel
lamellae with the orientation and wavenumber varying slowly on the
length-scale of the patterns (although we do not assume a global
orientation).  The dynamics of the lamellae are described by the phase
diffusion formalism \cite{POMEAU,CROSS}.  This formalism allows the
classification of the stability of the lamellae as well as a rigourous
criteria for wavenumber selection.  In this section we obtain the phase
diffusion equations for the laser induced melting model and discuss the
stability of the slowly varying lamellae state.  Using this description
we obtain the characteristic length-scale in agreement with the block
co-polymer length-scale and discuss a new instability which we conjecture
to be a signal of time-dependent chaotic dynamics.

Since we are interested in behaviour on length-scales much larger than
the wavelength of the stripes we need to separate the long distance,
large time behaviour on this length-scale from the short length-scale
behaviour.  To see how this is done, note that at each point in the
stripe pattern we can define the local wave vector ${\bf k}({\bf r},
t)$ which is directed in the direction normal to the stripes and whose
magnitude is the local wavenumber of the stripes.  Locally the choice of
this vector is two fold degenerate (i.e., we can choose either ${\bf k}$
or $-{\bf k}$).  However, once this choice is made at one point it is
fixed at all points.  We assume that ${\bf k}$ varies on length-scales
$1/\tilde{\epsilon}$, much larger than the pattern wavelength $2 \pi/k$.
(Here we use $\tilde{\epsilon}$ as the small parameter to avoid any
confusion with the previous $\epsilon$.)  We can now introduce the slow
space and time variables via
\begin{equation}
	{\bf X} = \tilde{\epsilon} {\bf r}, \; \; \; T =
	\tilde{\epsilon}^{2} t.
\end{equation}
and the fast phase variable, $\theta( {\bf r}, t)$,
such that $\theta = n \pi$ at each solid-liquid interface.  We can
relate the local wavevector ${\bf k}$ to the fast variable $\theta$
by
\begin{equation}
	{\bf k}( {\bf X}, T ) = \mbox{\boldmath{$\nabla$}} \theta(
	{\bf r}, t ).
\end{equation}
The goal of the phase diffusion description is to describe
the slow dynamics of ${\bf k}$ on the long length-scales ${\bf X}$.
The introduction of the separate slow and fast variables allows us to
write the dynamics order by order in $\tilde{\epsilon}$.
(The equilibrium free energy in terms of the phase variables has
been discussed by Kawasaki and Ohta \cite{KAWASAKI}.)

In terms of the phase variable the dynamical fields are of the form
\begin{equation}
	\phi( {\bf r}, t ) = \phi_{k}(\, \theta( {\bf r}, t ) \, ), \;
	\; u( {\bf r}, t ) = u_{k}( \, \theta( {\bf r}, t ) \, ),
\end{equation} where $\phi_{k}$ and $u_{k}$ are $2 \pi$-periodic functions
which are one dimensional time-independent solutions with wavenumber $k$.
Since static solutions exists for a band of wavenumbers, $\phi_{k}$
and $u_{k}$ depend on the local wavenumber and hence the slow variables
${\bf X}$ and $T$.  The full derivation is given in Appendix B.  Here we
only give the final results. The dynamics for the phase variable $\theta$
is
\begin{equation}
	\tau_{k} \; \partial_{t} \theta = -\mbox{\boldmath{$\nabla$}}
	\cdot {\bf k} \, G(k), \label{eq:PHASEDYNAMICS}
\end{equation}
where,
\begin{equation}
	G(k) = \frac{D}{r_{0}} A_{u}(k) q_{u,k} - A_{\phi}(k) q_{\phi,k},
\end{equation}
and
\begin{eqnarray}
	\tau_{k} & = & A_{\phi}(k) \left( q_{\phi,k} + \frac{\ell}{
	D k^{2} } \right) - \frac{1}{ r_{0} } A_{u}(k) q_{u,k}
		\nonumber \\
	& = & \frac{D-1}{ r_{0} } A_{u}(k) q_{u,k} + \frac{\ell}{ D k^{2}
	} A_{\phi}(k) - G(k).  \label{eq:TAUK}
\end{eqnarray}
The coefficients are related to the static one-dimensional
solutions by
\begin{eqnarray}
	A_{\phi}(k)
		& = &
	\frac{1}{ 2 \pi } \int^{2 \pi}_{0} d \theta \,
	\phi_{k}(\theta)^{2},
		\nonumber \\
	A_{\phi}(k) q_{\phi,k}
		& = &
	\frac{1}{ 2 \pi } \int^{2 \pi}_{0} d \theta \, \left(
	\partial_{\theta} \phi_{k}(\theta) \right)^{2},
		\nonumber \\
	A_{u}(k) & = & \frac{1}{ 2 \pi } \int^{2 \pi}_{0} d \theta
	\,u_{k}(\theta)^{2},
		\\
	A_{u}(k) q_{u,k} & = & \frac{1}{ 2 \pi } \int^{2 \pi}_{0}
	d \theta \, \left( \partial_{\theta} u_{k}(\theta) \right)^{2}.
		\nonumber
\end{eqnarray}
Since $\phi_{k}$ and $u_{k}$ depend on $k({\bf X},T)$
these coefficients depend on the slow variables.

The stability condition is apparent in curvilinear coordinates
\cite{CROSS}.  Let $w$ be the distance in the normal direction and ${\bf
s}$ be that in the tangential directions.  In curvilinear coordinates
the first order phase diffusion dynamics (Eq. (\ref{eq:PHASEDYNAMICS}))
becomes
\begin{equation}
	\tau_{k} \partial_{t} \theta
		=
	D_{w} \partial_{w}^{2} \theta + D_{s} \nabla^{2}_{s} \theta,
\end{equation}
where $D_{w} = -d (k G(k)) / dk$ and $D_{s} = -G(k)$.
The system is stable against slow variations of the local wavenumber (the
Eckhaus instability) if $\tau_{k}^{-1} D_{w} > 0$, while the lamellae is
stable against variations in the orientation (the zig-zag instability)
if $\tau_{k}^{-1} D_{s} > 0$.  If $\tau_{k} > 0$ and one neglects defects,
the functional \cite{CROSS},
\begin{equation}
	F\{ k( {\bf r}, t ) \} = \int d {\bf r} \int^{k( {\bf r},
	t )^{2}}_{0} d q^{2} G(q),
\end{equation}
defines a Liapunov functional with the extremum $G(k^{*})
= -D_{\parallel} = 0$.	Therefore the selected length-scale is exactly
the one marginally stable against the zig-zag instability.  Since, for
ordered lamellae, this wavelength is unique it must be the same as the
block co-polymer length-scale.

For some parameters, $\tau_{k}$ can be negative.  If this occurs
at larger wavenumber than that at which $D_{s}$ and $D_{w}$ becomes
negative, the lamellae will become unstable to both the zig-zag and
Eckhaus instabilities simultaneously.  That is both, $\tau_{k}^{-1}
D_{s}(k)$ and $\tau_{k}^{-1} D_{w}(k)$ becomes negative at the same $k$.
{}From Eq.\ (\ref{eq:TAUK}) this simultaneous instability supersedes the
isolated zig-zag instability if the latent heat is sufficiently small,
\begin{equation}
	\ell < (1-D) \frac{ D k^{*2} A_{u}(k^{*}) q_{u,k^{*}} }{
				r_{0} A_{\phi}(k^{*}) },
\end{equation}
where $k^{*}$ is the wavenumber at which the zig-zag
instability occurs, i.e., $D_{s}(k^{*}) = 0$.  Small $\ell$ is
exactly the condition for which the analogy with the relaxational
block co-polymer dynamics does not hold.  From the phase diffusion
dynamics, we cannot determine the final evolution of the unstable state.
However, the lamellae are unstable to perturbations of wavenumber $q$
in an entire band of wavenumbers around $q = 0$. So we conjecture that
this may be a signal of time-dependent dynamics.  Note that this is
not the skewed-varicose instability \cite{BUSSEandCLEVER} which has
also been suggested as a signal of a time dependent asymptotic state
\cite{GOLLUB,GREENSIDEandCROSS}.  The skewed-varicose instability is also
a long wavelength instability to perturbations with components both normal
and tangential to the stripes.	However, the skewed varicose instability
occurs at a fixed angle, i.e., fixed ratio of the normal and tangential
components of the perturbation.  In this case, the lamellae is unstable to
perturbations with an arbitrary ratio of normal and tangential components.

In general, the coefficients $A_{\phi}(k)$, $A_{u}(k)$ and $q_{\phi,k}$,
$q_{u,k}$ must be obtained numerically.  However, they can be obtained
explicitly in the strong and weak segregation limits.  In particular,
since these coefficients depend only on the static 1-d solutions we can
use the results of Appendix A.	(It turns out to be calculationally
simpler to obtain the selected wavelength using this criteria rather
than minimising the free energy.)

\subsubsection{Strong segregation limit}

In the thin interface limit, we can use the static solution in Appendix A,
to calculate the coefficients $A_{\phi}(k)$, $q_{\phi,k}$, $A_{u}(k)$
and $q_{u,k}$ order by order in ${\cal O}(\lambda^{-1}) = {\cal O}(
(r_{0}/D)^{1/3} )$.  Details are given in Appendix B.  We find that
\begin{equation}
	G(k) = -D_{s} = \frac{r_{0}}{D} \frac{\pi^{2} }{12  k^{4} }
	- \frac{1}{60} \left( \frac{r_{0}}{D} \right)^{2} \frac{\chi
	\pi^{4}}{ k^{6} } - \frac{\sigma}{\pi k}
			+ {\cal O}(\lambda^{-1}),
\end{equation}
where $\chi^{-1} = d \mu_{B} / d \phi |_{\phi=1}$. Here we
have kept the leading and next leading order terms. The diffusion constant
in the normal direction is
\begin{equation}
	\frac{ d}{d k }  k G(k) = -D_{w}
		=
	- \frac{ \pi^{2} r_{0} }{ 4 D k^{4} } + \frac{1}{6} \left(
	\frac{r_{0}}{D} \right)^{2} \frac{\chi \pi^{4}}{ k^{6} } +
	{\cal O}(\lambda^{-1}),
\end{equation}
The leading order $D_{w} > 0$, and the system is stable to
the Eckhaus instability at that order.	The kinetic coefficient $\tau_{k}$
is \begin{eqnarray}
	\tau_{k} & = &
	 \left( \frac{D - 1}{D} \right) \frac{\sigma}{\pi k} +
	\frac{\ell}{ D k^{2} } - G(k) + {\cal O}(\lambda^{-1}).
\end{eqnarray} The lamellar is unstable to the zig-zag instability ($G(k)
< 0$) for $\lambda > \lambda^{*}$ where \begin{equation}
	\lambda^{*} = \frac{2 \pi}{ k^{*} } = \left( \frac{ 96 D
	\bar{\sigma} }{ r_{0} }
		\right)^{1/3} + \frac{ 8 \chi \bar{\sigma} }{ 5 } +
		{\cal O}(\lambda^{-1}).
	\label{eq:LAMBDAzig}
\end{equation} This length-scale is in agreement with the  result in the
previous section.  Since $d k G(k)/dk < 0$ for all regions where $G(k)
< 0$, the lamellae is effectively stable against a separate Eckhaus
instabilities.

We now relate our result to the interfacial stability analysis of
Jackson and Kurtze \cite{JACKSON}.  They considered the stability
of a periodic array of stripes using a phenomenological interfacial
description, i.e., the kinetics of the temperature field was given by Eq.\
(\ref{eq:TEMPDYNAMICS}) and its value at the interface was fixed by the
local equilibrium  Gibbs-Thomson condition. The interfacial velocity was
related to the energy current by conservation of energy.  They discussed
the dynamics of an infinitesimal perturbation of wavenumber $q$ in the
orientation of the stripes.  Due to the undercooling (superheating) the
stripes can be unstable to the Mullins-Sekerka instability \cite{MULLINS}.
However, for low $q$, they found that the interface is stable because
the energy flux is decreased (increased) if the liquid phase advances
(recedes) into the solid phase.  At large $q$ the interface is stable due
to surface tension. The two stable bands  overlap only if the wavelength
of the stripes lies within a specific range.  Therefore the selected
wavelength must be within this stable band.  The largest wavelength in the
stable band was a factor of $(27/6)^{1/3}$ larger than that given by Eq.\
(\ref{eq:LAMBDAzig}).

In our analysis we also find a stable range of wavenumbers but we
explicitly showed that the selected wavelength is exactly the largest
wavelength in the band, i.e., the wavelength marginally unstable to
the zig-zag instability.  Moreover, we do not find the stability to
perturbations of low $q$.  The reason for this difference is that Jackson
and Kurtze discussed the stability of a single stripe independent of the
rest of the system.  Here we have considered collective behaviour of an
entire set of stripes.	Their analysis can be obtained from ours if we
consider a perturbation with a component in the normal direction $q_{w}=
{\cal O}(1/\lambda$).  Since the normal diffusion constant, $D_{w}$,
is positive, this means the interface is stable to perturbations of
this type at small tangential wavenumber $q_{s}$.  Furthermore due to
this stabilising effect, the stripe wavelength at which the instability
occurs is larger than that of the zig-zag instability.	In fact, Jackson
and Kurtze commented that preliminary estimates including cooperative
behaviour gives a largest stable wavelength approximately 2/3 that of
the independent stripe analysis \cite{JACKSON}.

The lamellae can become unstable to both the Eckhaus and zig-zag
instability simultaneously if $\tau_{k}$ becomes negative while both
$D_{s}$ and $D_{w}$ are positive.  We find this instability supersedes
(i.e., occurs at higher wavenumber) the independent zig-zag instability
if \begin{equation}
	\ell < (1-D) \frac{ \bar{\sigma} }{ \pi } k^{*} = (1-D) \left(
	\frac{ \bar{\sigma}^{2} r_{0}  }{ 12 D }
		\right)^{1/3}.
\end{equation}

\subsubsection{Weak segregation limit}

We can also obtain the phase equations in the weak segregation limit.
As shown in Appendix A, the static solution can be obtained order by
order in $\epsilon$ where $\epsilon^{2} = 1/4 - B$.  The coefficients
are given in Appendix B.  The result is \begin{equation}
	G(k) = \frac{\epsilon^{2}  A_{k}^{2} }{2} \left( 1 -
	\frac{r_{0}}{D k^{4} } + \epsilon^{4} \left( \frac{3  A_{k}^{2}
	}{128} \right)^{2} \left( 81 - \frac{r_{0}}{D k^{4} } \right)
	\right) + {\cal O}(\epsilon^{8}).
\end{equation} where $A_{k}^{2} = (8/3) ( 1 - 2 ( (k-k_{0})/\epsilon
)^{2} )$.  The lamellae becomes unstable to the zig-zag instability
at \begin{equation}
	k^{*4} = B  - \frac{5}{64} \epsilon^{4}
		+ {\cal O}(\epsilon^{6} ),
\end{equation} which is in agreement with the result for the equilibrium
wavelength.  $G(k)$ can also vanish if $A(k) = 0$.  This gives an upper
$k$ limit to the stability band which occurs at \begin{equation}
	k^{*} = k_{0} + \frac{\epsilon}{\sqrt{2}} + {\cal
	O}(\epsilon^{2}),
\end{equation} Here we have only kept the leading order.  To this
order this $k^{*}$ is equivalent to the neutral stability curve.
This wavenumber corresponds to a local maxima in the Liapunov functional.

To find the simultaneous Eckhaus/zig-zag instability we calculate
$\tau_{k}$ to leading order \begin{equation}
	\tau_{k} = \epsilon^{2}  A_{k}^{2} \frac{ r_{0} }{D } \frac{D-1}{2
	D k^{4} } + \epsilon^{2} A_{k}^{2} \frac{ \ell }{2 D k^{2} } -
	G(k)  + {\cal O}(\epsilon^{6}).
\end{equation} Therefore in the weak segregation limit the system becomes
unstable to both Eckhaus and zig-zag instabilities simultaneously if
$\ell < ( 1 - D ) \, (r_{0}/D)^{1/2} $.

We summarise the results of the stability analysis in Fig.\
\ref{fig:LENGTHS} in the large $\ell$ limit.  We show the neutral
stability curve $| k - k_{0} | = \sqrt{r_{0}/D}$ as well as the boundaries
of the zig-zag instability (and therefore the selected length-scale)
as determined by the expansions around the weak and strong segregation
limits.

\section{Numerical Results}

\subsection{Previous numerical works}

In this section we discuss the evolution of our model from an
initial random small amplitude state. Previous related studies
included numerical quenches of Raleigh-B\'{e}nard convection
\cite{MANNEVILLE83,GREENSIDE,ELDER} and block co-polymers
\cite{OONO,BAHIANA}.

Manneville \cite{MANNEVILLE83} integrated the Swift-Hohenberg equation
\cite{SWIFT} and other two-dimensional models of Raleigh-B\'{e}nard
convection starting from a small amplitude initial state.  To mimic
experiments the study was performed in a circular geometry.
He characterised the linear and weakly nonlinear regimes and studied
qualitatively the behaviour of defects.  Greenside and Coughran
\cite{GREENSIDE} were able to study a larger system and therefore make
a more extensive analysis.  They found that, near onset, the defects
evolved to the boundaries leaving a defect-free region in the middle.
However, further from onset, defects in the system tended to freeze
and the boundaries become less important. The description of parallel
roll regions separated by defects become less definite.  The qualitative
difference in the defect behaviour was also shown by studying specially
prepared initial conditions.  For a pair of disinclinations, they found
that, near onset, the disinclinations annihilate while far from onset,
the disinclinations persists forever \cite{GREENSIDE}.

More recently, Elder et al.\ studied the evolution of rolls
following a quench of the Swift-Hohenberg equation with a stochastic
noise term \cite{ELDER}.  They found that the system can undergo a
Kosterlitz-Thouless type transition as the noise strength is varied.
That is, for low noise amplitude, the orientation of the rolls exhibit
long range order, while for large noise, the orientational correlation
length is finite. They found that the width of the scattering intensity
peak decreases as $t^{-1/4}$ in time implying that regions of correlated
stripes grow as $t^{1/4}$.  Using a projection method to describe the
roll motion they found a small but non-vanishing coefficient for $D_{s}$
indicating a transition to $t^{1/2}$ growth at very late times, though
giving an effective  $t^{1/4}$ growth at intermediate times \cite{ELDER}.
However, the phase diffusion equations are the same generic form for
the Swift-Hohenberg equations \cite{CROSS,MANNEVILLE} as for our model.
The selected wavenumber will therefore be that marginally stable against
the zig-zag instability, i.e., the $k$ at which $D_{s}$ vanishes.
Based on the phase diffusion description we do not expect a crossover
to $t^{1/2}$ growth at late times.

Oono and Bahiana considered the behaviour of the dynamical model of block
copolymers (Eq.\ (\ref{eq:DYNAMICCOPOLYMER})) following a quench from an
initially disordered state \cite{OONO,BAHIANA}. They found a disordered
lamellae pattern and show that neither including a bending modulus nor
thermal fluctuations was sufficient to straighten out the patterns,
although straightening can occur due to boundary conditions and due to
the presence of flow \cite{BAHIANA}.  Numerically they were only able to
study not too small $B$ and found $\lambda \sim B^{-1/4}$. However, using
an analogy with phase ordering they argued that $\lambda \sim B^{-1/3}$
for small $B$.	Liu and Goldenfeld also constructed a scaling function
\cite{LIU} to describe the behaviour of the characteristic length-scale as
a function of $B$ and time $t$ after the quench.  This scaling hypothesis
was partially confirmed \cite{BAHIANA}.

\subsection{Selected length-scale}

We numerically updated the coupled partial differential equations
(\ref{eq:PHIDYNAMICS}) and (\ref{eq:UDYNAMICS}) using an Euler
discretization with mesh-size $dx = 1.25$, time-step $dt = 0.2$ and a
sphericalized Laplacian (see Appendix C for details).  We used $512 \times
512$ lattices with smaller lattices to test for finite size effects.
Our simulations were restricted to the symmetric case ($\Delta j = 0$)
and fixed $D = 0.5$.  For the initial set of simulations we fixed $\ell
= 2$.  This value of $\ell$ was sufficiently large so that the copolymer
analogy is applicable and a static state is always reached.  Most of the
runs were to $t = 102400$ or about 500000 updates.  For selected cases,
we ran to $t = 256000$.  We chose an uncorrelated Gaussian distribution
for the initial values of $\phi$ with mean zero and deviation of $0.001$.
The $u$ field was set to zero initially.  If we start with an average
$u$ different from zero, the system rapidly evolved so that the average
was zero .  Even starting with a different value of average $\phi$
did not have a major effect unless the value was outside the spinodal.
Unfortunately due to the length and size of the simulation we were
only able to average over two to four initial conditions depending on
the parameters.

In Fig.\ \ref{fig:PICTURE} we showed that there are two regimes displaying
qualitatively different behaviour.  Near onset, the system evolves to
locally parallel rolls with the wavenumber fixed near that maximising
the linear instability i.e., $k_{0} = 1/\sqrt{2}$.  The parallel rolls
then slowly straighten out in time.  As shown in Fig.\ \ref{fig:PROFILE}a,
$\phi$ and $u$ are relatively small amplitude and approximately sinusoidal
justifying our near onset analysis.  As expected $u$ is negative
(positive) where $\phi$ is positive (negative) so  that the solid
phase ($\phi < 0$) is superheated and the liquid phase ($\phi > 0$) is
undercooled.  Far from onset (Fig.\ \ref{fig:PROFILE}b) the characteristic
wavenumber slowly decreases with time and is much less than $k_{0}$.
There does not seem to be any tendency toward locally parallel rolls
and the system freezes into a disordered lamellar structure. The solid
phase is superheated and liquid phase undercooled. The $\phi$ field can
be divided into bulk and interfacial areas where $\phi$ changes rapidly
from one bulk value to another while $u$ is smooth.

Figure \ref{fig:SCATTERING} shows the circular average scattering
intensity $S_{k} = \langle \; \phi_{k} \phi_{-k} \; \rangle$ for the
order parameter field for the same values of $r_{0}/D$.  Near onset, the
wavenumber of the very sharp peak is fixed in time.   Far from onset,
we observe a much broader peak which moves, with time, toward lower
wavenumber.  This is indicative of the growing characteristic wavelength
of the patterns and the disordered structure observed far from onset.
Similar scattering intensities have been observed experimentally by
Dworschak et al. \cite{DWORSCHAK}.

As a quantitative measure of the characteristic wavenumber we chose
a quantity that amplifies the contribution from the peak in $S_{k}$.
For each value of $r_{0}/D$ we calculated \begin{equation}
	k_{1} = \frac{ \int^{\infty}_{0} dk \, k \, S_{k}^{2} }{
	\int^{\infty}_{0} dk \, S_{k}^{2} }.
\end{equation} The squares in Fig.\ \ref{fig:LENGTHS} show $k_{1}$
vs.\ $r_{0}/D$ for various values of $r_{0}$.  Also shown are the
selected wavelengths obtained from the stability analysis/ free energy
minimisation.  We show the lowest order and next order result from the
far from onset expansion but only the lowest order result for the near
onset expansion since the next order correction is negligible where this
is valid. There is very good agreement between the theoretical prediction
and $k_{1}$ in the simulation.	In particular, we observe both limits
of the strong segregation regime in which $k \sim (r_{0}/D)^{1/3}$
and the weak segregation regime for which $k \sim (r_{0}/D)^{1/4}$.

Liu and Goldenfeld have conjectured that the characteristic wavenumber
has the following scaling form \cite{LIU} \begin{equation}
	k( t, B )
		=
	k_{\infty}( B ) g( t B ), \label{eq:SCALINGK}
\end{equation} where, using the block co-polymer notation, $B = r_{0}/D$
and $g(x)$ is a dimensionless scaling function that behaves as $x^{-1/3}$
for small $x$.	This conjecture was based on dimensional analysis of the
block co-polymer equations and the analogy with phase ordering dynamics.
They numerically tested this by plotting $k(t,B)/k_{\infty}$ vs. $t B$
and found a good collapse \cite{LIU,BAHIANA}.  However, for small $B$
they were not able to obtain the asymptotic wavenumber $k_{\infty}$
directly and they chose $k_{1}(\infty)$ to produce the best collapse.
Therefore it is not entirely clear that Eq.\ (\ref{eq:SCALINGK}) is
obeyed.  In fact, it is simple to see that this scaling form cannot hold
for all $B$ and $t$.  For $B$ near onset $k(t)$ is approximately $k_{0} =
\sqrt{2}$ from the very earliest to latest times.  Therefore near onset,
$g(x)$ is a constant for all $x$.  Furthermore the timescales over which
linear dynamics hold, $t_{lin}$, can be made arbitrary large by decreasing
the amplitude of the initial condition. There must be an offset in time
which depends on $t_{lin}$.

Therefore this scaling form can hold only if we impose the constraint that
$t \gg t_{lin}$ and $k_{1} < k_{0}$.  In Fig.\ \ref{fig:SCALING} we plot
$k_{1}(t)/k_{1}(\infty)$ vs. $t r_{0}/D$ for values of $r_{0}/D$ ranging
from 0.00046875 and 0.24 where $k_{1}(\infty)$ are the values shown in
Fig.\ \ref{fig:LENGTHS}.  We find a  reasonable collapse of data over this
regime although, due to the small number of configurations, the scatter is
quite large. In the strong segregation limit, one expects that the small
value of $r_{0}/D$ is irrelevant at early times and we recover the result
for spinodal decomposition with conservation of order parameter,, i.e.,
$k_{1}(t) \sim t^{-1/3}$ or $g(x) \sim x^{1/3}$ \cite{LIU}.  Therefore,
with the restrictions discussed, which essentially means that the scaling
relation is only nontrivial not too close to onset, our data is consistent
with the scaling form of Liu and Goldenfeld \cite{LIU}.

\subsection{Orientation correlations}

To further quantify the difference in the strong and weak segregation
regimes we construct the director field for the stripes.
The stripes pattern can be describe
by a liquid crystal type order parameter \cite{TONER}.
To construct the director field, we cannot use
the local gradient since it will vanish at maximas and minimas of the
stripe pattern.  Instead we use a symmetric difference to determine
the direction tangential to the stripes.  We define at each lattice
point $i, j$ the differences
\begin{eqnarray}
	\Delta(0) & = & | u_{i,j} - u_{i+1,j} | +
		| u_{i,j} - u_{i-1,j} |,
		\nonumber \\
	\Delta( \pi/4 ) & = & \frac{1}{\sqrt{2} }
		\Bigr( | u_{i,j} - u_{i+1,j+1} |
		+ | u_{i,j} - u_{i-1,j-1} | \Bigr),
		\nonumber \\
	\Delta( \pi/2 ) & = & | u_{i,j} - u_{i,j+1} | + | u_{i,j} - u_{i,j-1} |,
		\\
	\Delta( 3 \pi/ 4) & = &
	\frac{1}{\sqrt{2}}
		\Bigr( | u_{i,j} - u_{i+1,j-1} |
		+ |  u_{i,j} - u_{i-1,j+1} | \Bigr).
	\nonumber
\end{eqnarray}
We then define $\theta$ as the angle minimising $\Delta(\theta)$.  This is
obtained by finding the $m^{*}$ minimising $\Delta( m \pi/4 )$ and making
a quadratic fit to $\Delta( (m^{*}+1) \pi/4 )$, $\Delta( m* \pi/4 )$
and $\Delta( (m^{*}-1) \pi/ 4)$.  Hence $\theta$ is in the direction in
which the change in $|u|$ is the smallest (tangential to the stripes).
We use $u$, since for $r_{0}/D$, $\phi$ is effectively discontinuous
on length-scales larger than the interfacial width.

The orientational correlation function measures
the changes in orientation of the stripes.  We define it as
\begin{equation}
	C_{\theta}( {\bf r} )
	= \langle \; \exp\left( i 2 \theta( {\bf R} + {\bf r} ) \right)
		\exp\left( i 2 \theta( {\bf r} ) \right) \; \rangle,
\end{equation}
where the angular brackets indicate an average over initial conditions or,
for sufficiently large systems, a translational average over ${\bf R}$.
The factor of two is required since the director
is a headless vector and has a two-fold symmetry.

The qualitative difference between the behaviour
close to and far from onset is magnified in the director field.
Figure \ref{fig:OPICTURE} shows the director field for $r_{0}/D =
0.24$ for the same times as in Fig. \ref{fig:PICTURE}.  At $t = 1600$,
the system is barely out of the linear dynamics stage and one observes
essentially randomly oriented regions.  Patches of correlated regions
form and grow with time. By $t=12800$ the correlated regions are already
on the order of the system size.  The patches continue to grow but there
are long lived structures corresponding to grain boundaries
and defects.

Figure \ref{fig:ORIENTATION}a shows $C_{\theta}(x)$ near onset, $r_{0}/D
= 0.24$ and $t=$ 800, 1600, 3200, 4800, 6400 and 12800.  $C_{\theta}$
coincide for the first two times since they are just beyond the linear
regime.  For $t \geq 12800$ the orientational correlation length is very
large and strong finite size effects are found.  As shown in the inset,
there are two characteristic length-scales evident in $C_{\theta}$.
There is a kink in $C_{\theta}$ at approximately $x \sim 3$, and then
a slow decay at larger $x$.  We interpret this kink as the contribution
arising from the defect cores while the slow decay comes from the slow
variations in the orientation.  With increasing time the contribution from
the `defects' decreases reflecting the decrease in the density of defects.
This is consistent with our qualitative analysis of the patterns.

Figure \ref{fig:ORIENTATION}b shows $C_{\theta}(r)$ far from onset,
$r_{0}/D = 0.001875$.  Times $t=800, 3200, 12800$ and $102400$ are
shown.  In contrast to the near onset case, there is no kink at short
length-scales since pattern can no longer be described by regions of
locally parallel lamellae separated by defects.  The orientational
correlational length is much smaller than for that near onset and
grows slowly in time.  This growth is due mainly to
the growth of the characteristic pattern size during this time.

To quantify the behaviour further, we define the orientational correlation
length $L_{\theta}(t)$ as the value of $x$ at which $C_{\theta}(x)$
reaches the value of $1/e$ (Note that $C_{\theta}(0) = 1$).  Figure
\ref{fig:LTHETAONSET} shows $L_{\theta}(t)$ for $r_{0}/D = 0.06, 0.12,
0.18$ and $0.24$.  Very close to onset ( $r_{0}/D = 0.18$ and $0.24$),
$L_{\theta}$ is consistent with the power law growth $L_{\theta} \sim
t^{1/4}$ as observed by Elder et al.\ \cite{ELDER} in a quench of the
Swift-Hohenberg equation.  However, with decreasing $r_{0}/D$ we find
that $L_{\theta}$ saturates at a value which depends on $r_{0}/D$ and
is much smaller than the system size, although asymptotic logarithmic
growth cannot be ruled out.

Figure \ref{fig:LTHETAK} is a plot of asymptotic value of
the product $L_{\theta} k_{1}$ as
a function of $r_{0}/D$. For the two values of $r_{0}/D$ near onset
$L_{\theta}$
continues to grow  so that the points are not the saturation
values.  Several features are evident.  The product $L_{\theta}
k_{1}$ approaches a well define limit of order unity as $r_{0}/D$
approaches zero.  This is the thin interface regime in which there
is single length scaling with the orientational correlation length
being on the same order as the characteristic wavelength of the patterns.
Closer to onset, the system cannot be described by a single length-scale
but instead both the characteristic pattern size and the orientational
correlational length is required.  The orientational correlation length
becomes much larger than the pattern size as $r_{0}/D \rightarrow 1/4$.

Several interpretations are possible.  First, $L_{\theta}(t)$ may
grow as a power law $L_{\theta}(t) \sim t^{1/4}$ up to some finite
saturation value depending on $r_{0}/D$.  Candidates for this saturation
length are the length-scales on which defects and boundary effects heal
\cite{CROSSdefect}.  Near onset, the perturbation due to a defect will
disappear on length-scales of $\epsilon^{-1}$ in the direction normal
to the stripes and $\epsilon^{-1/2}$ in the direction tangential to
the stripes ($\epsilon^{2} = 1/4 - r_{0}/D$ ).  The distinction in
the two directions is due to the vanishing of the tangential diffusion
constant at the selected wavenumber.  Estimates of the first length-scale
gives $\approx 10$ for $r_{0}/D = 0.24$ and $3.7$ for $r_{0}/D = 0.18$.
Estimates of the tangential length-scale gives $\approx 3$ and $\approx
2$ respectively.  Although these candidates cannot be ruled out they are
much smaller than for the observed orientational correlation length.
These lengths are equivalent to the equilibrium or steady state
correlation length on which the effect of fluctuations decay \cite{TONER}.

A second possibility is that the orientational correlation length
grows as a power law in time for small $t$ and then crosses over to
logarithmic growth at larger time with the
cross-over length depending on $r_{0}/D$.  This slow growth will then
continue until limited by the system size. It would indicate that
the long time scale dynamics are similar to that of a magnetic system in
a random field, perhaps, because the long-lived defects are effectively
quenched-in impurities.  Based on our simulations we cannot
rule this out.  This crossover length-scale may be related to steady
state correlation length described above.

The third possibility is the most intriguing.  According to simulations
of the Swift-Hohenberg equation by Elder et al.\ \cite{ELDER}, the system
shows long range orientational correlational ordering at a finite distance
from onset.  That is, the orientational correlation length is limited
only by the system size.  On the other hand our results indicate that,
further from onset, the orientational correlation length is finite and
much smaller than the system size.  If this is the case then there will be
a value of $r_{0}/D < 1/4$ at which the orientational correlation length
becomes finite (in an infinite system).  This would mark the presence
of a nonequilibrium phase transition similar to a glass transition.
Note that this transition cannot be an equilibrium phase transition since
the lowest free energy state is that of parallel stripes.  In addition,
this transition, if it exists, must depend on dynamical features such
as quench rate, presence of walls and in a real system impurities.
Such a transition would depend on the quench rate since the
lamellae will straighten if the quench is slow enough \cite{CHAKRABARTI}.
Although such a possibility is intriguing, based on our numerical results,
we cannot differentiate between this scenario and other scenarios based on
an effective crossover length approaching infinity as $r_{0}/D \rightarrow
1/4$ in which case there is no secondary transition for $r_{0}/D < 1/4$.
To tell which of these scenarios, if any, occur one must simulate larger
systems with more initial configurations and with a systematic finite
size analysis.

\subsection{Time-dependent Asymptotic States}

In the limit of large latent heat  the laser induced melting model is
equivalent to a dynamical model of block copolymers which has a Liapunov
functional and, therefore, purely relaxational dynamics.  In this limit,
there can be no motion of the patterns in the asymptotic state.  However,
as the latent heat is decreased the dynamics of the laser induced melting
becomes distinctly different.

For illustrative purposes we discuss what happens
as $\ell$ is decreased for particular values of $r_{0} = 0.03$
and $D = 0.5$.  There is very little change in the pattern dynamics as
$\ell$ is decreased from 2 to 1.  We monitor $\bar{\phi}(t) = \sum_{i,j}
\phi_{i,j}(t) / (n_{x} n_{y})$, i.e., the average value of $\phi$, as a
function of time. As $\ell$ is decreased below unity there are
small oscillations of $\bar{\phi}$ at early times.  These oscillations
decay after one or two cycles and a disordered lamellar pattern is formed
similar to that for $\ell = 2$.  As $\ell$ is decreased further, below
$0.6$, there is a large increase in the amplitude of the oscillations as
well as in the time for which oscillations persist.  The corresponding
patterns during this transient period consists of regions of disordered
stripes intermingled with oscillating structures.  For $\ell < 0.5$,
there is a long transient period with a chaotic space time pattern.  This
transient dynamics persists for many periods but the system eventually
evolves to a state that is homogeneous in space but oscillating in time.

For $r_{0} = 0.03$ and $L_{x} = L_{y} = 160$, we can locate the transition
to within $\Delta \ell = 0.05$. We always find the stripe phase
for $\ell = 0.55$ and the oscillating homogeneous phase for $\ell = 0.5$.
Depending on initial conditions, states for $\ell$ between these two
values may settle into a inhomogeneous state consisting of
stripes and oscillating regions or may settle into a static stripe phase
or a oscillating homogeneous
state.  (The smaller the system the broader the range
of $\ell$ on which one finds mixed states.)  The lifetime of transient
can vary by one order of magnitude with different initial conditions
and is rapidly increasing function of system size.

A quantitative measure of the persistence time is the time required
for the system to evolve to a state in which $\phi_{i,j}$ is the
same sign for all $i,j$.  Defining $\log \tau = \langle \, \log \tau
\rangle$ where $\langle \bullet \rangle$ indicates the average over
the initial conditions, we find that for $r_{0} =0.03$ and
$\ell = 0.5$  $\tau \approx 2700$
for $L_{x}=L_{y} = 160$, $\tau \approx 650$ for $L_{x}=L_{y} = 80$ and
$\tau \approx 170$ for $L_{x} = L_{y} = 40$ (The averages are over 30-80
initial conditions.).  The lifetime of transient state grows rapidly with
system size and is consistent with diffusion being the dominant process
since $\tau \sim L_{x}^{2}$.  For infinite systems the transient state
may persist indefinitely.  (Note that we have taken the average of the
$\log(\tau)$ rather than $\tau$ itself so that rare long-lived states
do not dominate the average.)

Figure \ref{fig:CHTRANSITION} shows the value of $\ell$ at which the
transition from stripes to oscillating homogeneous states occur as a
function of $r_{0}$. We find $\ell^{*}_{1}$ and $\ell^{*}_{2}$ such
that for all $\ell < \ell^{*}_{1}$ the asymptotic state is a homogeneous
oscillating state and for all $\ell > \ell^{*}_{2}$ the asymptotic state
is a disordered stripe phase.  In this way we can bracket the transition
to within a narrow band of $\ell$.  In addition to $\ell^{*}$
decreasing with $r_{0}/D$ we also find that the period of the oscillations
decreases rapidly with $r_{0}/D$.  For example for $r_{0}/D = .03$ the
period is about 50 and for $r_{0}/D = .00375$ the period is about 500.

One possible reason for the oscillatory behaviour is that the
linear dispersion relation (Eq.\ (\ref{eq:LINEAR})) predicts an
oscillatory instability.  The linear stability analysis predicts a
complex $\omega_{k}$ for $k = 0$, if $ 1 - 2 \sqrt{r_{0}} < \ell <
1 + 2 \sqrt{r_{0}}$ and that $\omega_{k}$ to be unstable if $\ell
< 1$.  The band with complex $\omega_{k}$ is also shown in Fig.\
\ref{fig:CHTRANSITION}.  It is clear that we can rule out the oscillatory
linear dynamics as the reason for the oscillatory asymptotic state.  In fact,
except near onset, we do not see any reflection of the linear behaviour in
the final state.  For example, we do not see any qualitative differences
as we cross $\ell = 1$ even though this is an important boundary in the
linear analysis as the $k=0$ mode is unstable for $\ell < 1$ and stable
for $\ell > 1$.

We also plot the values of $\ell$ at which the simultaneous
Eckhaus/zig-zag instability occurs.  Our data is primarily for the weak
segregation case in which $\ell^{*} \sim (r_{0}/D)^{1/2}$.  We find
that, the $\ell$ for the Eckhaus/zig-zag instability is approximately
a factor of four smaller than $\ell^{*}$, the value of $\ell$ at which
the transition occurs in our numerical experiment.  However, the scaling
behaviour seems to agree in this range of parameters and the result is
consistent with the requirement that $\ell^{*}$ is larger than that
predicted by this instability.

\section{Summary}

To summarize, we have introduced a continuum model of laser induced
melting.  In this model, the presence of coexisting solid-liquid regions
is due to the higher reflectivity of the
molten regions.   We showed that,
in the limit of large latent heat, the dynamics becomes equivalent to a
model of block copolymers \cite{OONO,LIU,BAHIANA}.
In particular, any static state of the laser induced melting model is also a
static state of the block co-polymer model.  As the latter is based on an
equilibrium free energy this analogy is used to obtain the wavenumber
of the patterns observed in the laser induced melting experiments.
Via this analogy we justified the approximate free energy argument of
Hawkins and Biegelsen \cite{HAWKINS}.

We also derived the phase diffusion equations describing the slowly
varying stripes.  The selected wavenumber, $k$, of the patterns is the one
in which the tangential diffusion constant, $D_{s}(k)$, vanishes, i.e.,
the wavenumber marginally stable to slow variations in the orientation
(the zig-zag instability).  Therefore the wavenumber of the lamellae
is also given by the stability criteria.  We obtained this wavenumber
via systematic expansions from two limits, near onset (weak segregation)
and far from onset (strong segregation).  The wavenumber is
in good agreement with experiment corresponding to the strong segregation
limit \cite{DWORSCHAK}. These are very general results for a
wide class of systems.  In particular, we argue that the vanishing of the
tangential diffusion constant at the selected wavenumber is the reason
that the length-scale on which regions are orientationally correlated
grow as $t^{1/4}$ in simulations of quenches of the Swift-Hohenberg
equation \cite{ELDER}.

We also discussed a new instability. The lamellae can become unstable to
the zig-zag instability and Eckhaus instability (slow variations in the
wavenumber) simultaneously.  We conjecture that this instability is a
signal of time-dependent asymptotic states which are observed experimentally.

We numerically study the behaviour after a quench.   We find strong
agreement between our analytic prediction for the selected wavenumber,
and the numerical simulation results as indicated by the peak in the
scattering intensity.  There are two qualitatively different
patterns determined by the distance (in parameter space) from onset.
Near onset, locally parallel stripes slowly straighten out
in time.  Far from onset, a disordered lamellar structure is formed.
This qualitative difference is also reflected in the
scattering intensity.

We constructed the director field and measure the orientational
correlation length.  Close to onset, the behaviour of the orientational
correlation length is consistent with a $t^{1/4}$ growth with the final
length fixed by the system size.  Further from onset the
correlation length quickly saturates.  The product of the orientational
correlation length and the selected wavenumber approaches a limit
with decreasing $r_{0}/D$ (i.e., far from onset).  Based on our data
and the conclusions of a simulation of the Swift-Hohenberg equation by
Elder et al., \cite{ELDER}, we conjecture that there may be a dynamical
phase transition dividing a regime near onset where the orientational
correlation length grows until limited by the system size and a second
regime in which the orientational correlation length saturates at a value
which depends on  the distance from onset.  Much more extensive numerical
work will be required to determine the validity of this conjecture.

Finally we studied dynamics for smaller latent heat in which the system
may evolve to a time-dependent asymptotic state.  In this case we observe
a long-lived transient state which eventually evolves into a homogeneous
oscillating state.  The persistence time of the transient state depends on
the system size.  We sketch out the boundary between this and the regime
in which the static steady state is reached as a function of latent heat
$\ell$ and reflectivity difference $r_{0}$.  We find that the value of
$\ell$ at which this transition occurs is larger than that predicted
by the simultaneous Eckhaus/zig-zag instability.  However, the scaling
behaviour for the phase boundary agrees with that of this instability.
This indicates that this instability marks the limit of metastability
of the locally parallel stripes.

\acknowledgements
We are grateful to H.\ van Driel, D.\ Kurtze, C.\ Sagui, T.R.\ Rogers
and Y.\ Oono for helpful comments and discussions.  We are grateful
to D.\ Jasnow for allowing us to clog up his workstation.  This work
was supported by the Natural Sciences and Engineering Research Council
of Canada.

\appendix

\section{Static Solutions}

In this appendix we show that the static solutions can be obtained via
systematic expansions around the small amplitude and sharp interface
limits.  Assuming periodic stripes of wavenumber $k$ and the normal in
the $x$ direction, the one-dimensional static solution obeys
\begin{equation}
	0 = \partial_{x}^{2} \left( \mu_{B}( \phi ) +
	\partial_{x}^{2} \phi \right) - B \phi.
			\label{eq:BSTATICCOPOLYMER}
\end{equation}
The reduced temperature field $u$ can be obtained using the relation
$\partial_{x}^{2} u = r_{0} \, \phi/D$.

\subsection{Sharp Interfacial Limit}

For small $B$, we break up the system into four sections consisting
of two outer (bulk) regions corresponding to the positive and negative
$\phi$ phases and two inner regions consisting skin of thickness unity
around the two interfaces.  In the outer regions we are interested in
length-scales of order $\lambda \sim k^{-1}$.  We introduce the small
parameter $\epsilon = \lambda^{-1}$.  To extract the large distance
behaviour, we rescale the normal direction by $X$ = $\epsilon x$
in the outer region.  The interfacial width is order unity so the
normal direction is not rescaled in the inner region.  $B$ must also
be rescaled since it determines the pattern wavenumber.  In the bulk
phases $\phi = \pm 1$, as $B \rightarrow 0$.  Using this condition
in Eq.\ (\ref{eq:BSTATICCOPOLYMER}) gives $B = {\cal O}(\epsilon^{3})$.
Therefore we rescale $B$ by $\tilde{B} = \epsilon^{-3} B$.

In the outer regions the static solution obeys
\begin{eqnarray}
	0
	& = &
	\partial^{2}_{X} \mu(\phi)
	 - \epsilon \tilde{B} \; \phi.
	\label{eq:STATICOUTER}
\end{eqnarray}
where $\mu(\phi) = \mu_{B}(\phi) + \epsilon^{2} \partial^{2}_{X} \phi$.
In the inner region the static solution is given by
\begin{eqnarray}
	0
	& = &
	\partial^{2}_{x} \mu(\phi)
	 - \epsilon^{3} \tilde{B} \; \phi,
	\label{eq:STATICINNER}
\end{eqnarray}
where $\mu(\phi) = \mu_{B}(\phi) + \partial^{2}_{x} \phi$.
In each region we expand $\phi = \phi_{0} + \epsilon \phi_{1} +
{\cal O}(\epsilon^{2})$ and $\mu(\phi) = \mu_{0} + \epsilon \mu_{1} +
{\cal O}(\epsilon^{2})$.  The matching conditions for $\mu$ at the
boundary between inner and outer domains are $\phi^{inner}_{i}
= \phi^{outer}_{i}$ and $\partial_{x}^{m} \phi^{inner}_{i+m} =
\partial_{X}^{m} \phi^{outer}_{i}$.

The inner equation gives
\begin{equation}
	\phi(x) = \phi_{0}(x) + {\cal O}(\epsilon^{2}),
\end{equation}
where $\phi_{0}$ is the planar interfacial profile for $B = 0$.  (We have
used the condition that $\phi_{i > 0}$ is orthogonal to the Goldstone
translation mode $d \phi_{0}/d x$.)  The outer equation gives
\begin{equation}
	\phi(X) = \pm \left( 1
		+ \epsilon
		\chi \tilde{B} X
	\left( X \mp \frac{1}{2} \right) \right)
		+ {\cal O}(\epsilon^{2}),
		\label{eq:PHISTRONG}
\end{equation}
where $\chi^{-1} = d \mu_{B} / d \phi |_{\phi=1}$ and $X = x/\lambda$.
For $\mu_{B} = -\phi + \phi^{3}$ and $\chi = 1/2$. In the outer region
the temperature field becomes
\begin{equation}
	u(X) =  \pm
		\left( \epsilon \frac{ \tilde{B} }{ 2 } X
		\left( X \mp \frac{1}{2} \right)
		+ \epsilon^{2} \chi \tilde{B}^{2} \frac{X}{12}
		\left( X^{2} \left( X \mp 1 \right) \pm \frac{1}{8} \right)
		\right)
		+ {\cal O}(\epsilon^{3}).
		\label{eq:USTRONG}
\end{equation}
In the inner region $u(x) = {\cal O}(\epsilon^{3})$.

\subsection{Weak Segregation Limit}

In the weak segregation limit we use standard methods
\cite{MANNEVILLE} to expand the solution around the critical point $B =
B_{0} = 1/4$, $k = k_{0} = 1/\sqrt{2}$.  We define a small parameter
$\epsilon$ defined by $B_{0} - B = \epsilon^{2}$ and assume $k = k_{0}
+ \epsilon k_{1}$. The static solution obeys
\begin{equation}
	0 = -\partial_{x}^{2} \left( \phi - \phi^{3} +
		\partial_{x}^{2} \phi \right)
		- (B_{0} - \epsilon^{2}) \phi,
\label{eq:WEAKSTATIC}
\end{equation}
where we have used $\mu_{B} = -\phi + \phi^{3}$.
We want a solution for wavelength $\lambda = 2 \pi/k$.
We can write the solution as
\begin{equation}
	\phi(x) = \sum_{m=1}^{\infty} \epsilon^{m} f_{m}(k x),
\end{equation}
where $f_{m}(\theta) = f_{m}( \theta + 2 \pi)$.  To fix the position
of the interface, we require that $f_{m}$ be orthogonal to $f_{m}'$.
We can now write Eq.\ (\ref{eq:WEAKSTATIC}) order by order in $\epsilon$.
First order gives
\begin{equation}
	0 = {\cal L}_{0} f_{1}(\theta),
\end{equation}
where ${\cal L}_{0} = -\left( k_{0}^{2} \partial_{\theta}^{2}
+ k_{0}^{4} \partial_{\theta}^{4} + B_{0} \right)$.
This gives $f_{1}(\theta) = A_{k} \sin( \theta )$. (We assume odd symmetry.)
The second order expression along with the orthogonality condition gives
$f_{2} = 0$.  The third order condition is
\begin{equation}
	-{\cal L}_{0} f_{3} =
		 -\left( k_{1}^{2} \partial_{\theta}^{2}
			+ 6 k_{1}^{2} k_{0}^{2} \partial^{4}_{\theta}
			- 1 \right) f_{1}
			+ k_{0}^{2} \partial_{\theta}^{2} f_{1}^{3}.
	\label{eq:THIRD}
\end{equation}
The solvability  condition requires that the RHS is orthogonal to the
zero eigenvector of ${\cal L}_{0}$ which is simply $\sin(\theta)$.
Applying this condition gives
\begin{equation}
	\frac{3}{8} A_{k}^{3} = k_{1}^{2}
	A_{k} - 6 k_{1}^{2} k_{0}^{2} A_{k} + A_{k}  =
	(1 - 2 k_{1}^{2}) A_{k},
\end{equation}
or $A_{k}^{2} = 8 ( 1 - 2 k_{1}^{2} )/3$.  Substituting this into Eq.\
(\ref{eq:THIRD}) and solving the resulting ODE gives $f_{3}(\theta) =
A_{3k} \sin( 3 \theta )$ with $ A_{3k} = 9 A_{k}^{3}/128$.  The next order
expression gives $f_{4}(\theta) = 0$.  Therefore the static solution is
given by
\begin{equation}
	\phi(x) = \epsilon A_{k} \sin( k x ) + \epsilon^{3}
		A_{3k} \sin( 3 k x) + {\cal O}(\epsilon^{5}),
	\label{eq:PHIWEAK}
\end{equation}
and
\begin{equation}
	u(x)  =
	\epsilon \frac{A_{k}}{ k^{2} } \sin( k x ) +
	\epsilon^{3}\frac{ A_{3 k }}{9 k^{2} }
	\sin(3 k x) + {\cal O}(\epsilon^{5}).
	\label{eq:UWEAK}
\end{equation}


\section{Phase Diffusion Equations}

\subsection{Rescaled variables and separation of length-scales}

In this appendix we derive the phase diffusion description of the laser
induced melting model.  Our derivation closely follows the derivation of
Cross for convection models \cite{CROSS}.

We assume that the system consists of stripes with slowly varying
wavenumber and orientation.   There are two relevant length-scales,
first, the wavelength of the patterns and second, the length-scale on
which the orientation and wavenumber varies.  We are interested in the
behaviour on this second, large length and long time scales.  In order to
separate the two length-scales we introduced the slow variables
\begin{equation}
	{\bf X} = \epsilon {\bf r}, \; \; T = \epsilon^{2} t,
\end{equation}
where $\epsilon$ is a small parameter and the rapidly varying phase
variable $\theta( {\bf r}, t)$.  This mixture of slow and fast variables
is somewhat cumbersome so we introduce a slow phase $\Theta = \epsilon
\theta$. The local wavevector ${\bf k}$ of the stripes is
\begin{equation}
	\mbox{\boldmath{$\nabla$}} \theta( {\bf r}, t ) =
	\mbox{\boldmath{$\nabla$}}_{\bf X} \Theta( {\bf r}, t ) =
		{\bf k}( {\bf X}, T ).
\end{equation}
The dynamical fields are of the form
\begin{equation}
	\phi( {\bf r}, t) = \phi(\theta( {\bf r}, t), {\bf X}, T ), \; \;
	u( {\bf r}, t) = u(\theta( {\bf r}, t), {\bf X}, T ),
\end{equation}
where $\phi$ and $u$ are $2 \pi$ periodic function of $\theta$ whose
functional form  depends on the local wavenumber $k$.

In the new coordinate systems, the gradient becomes
$\mbox{\boldmath{$\nabla$}}
= {\bf k} \partial_{ \theta }
+ \epsilon \mbox{\boldmath{$\nabla$}}_{{\bf X}}$.
The Laplacian is
\begin{equation}
	\nabla^{2} = k^{2} \partial_{\theta}^{2}
	+ \epsilon D_{1} \partial_{\theta} +
	\epsilon^{2} D_{2}
\end{equation}
where $D_{1}$ and $D_{2}$ depends only on the slow variables ${\bf X}, T,$
with
\begin{equation}
	D_{1} = 2  k^{2} \partial_{W} + k \frac{ \partial k}{ \partial W}
	 + k {\cal K}_ {W}
\end{equation}
and
\begin{equation}
	D_{2} = k {\cal K}_{W} \partial_{W} + \ell {\cal K}_{S} \partial_{S}
	+ \left( k \partial_{W} \right)^{2} +
	\left( \ell \partial_{S} \right)^{2}.
\end{equation}
Here $W$ is the rescaled distance in the normal direction with
${\bf \hat{W}} \cdot \mbox{\boldmath{$\nabla$}}_{\bf X} =
k \partial_{W}$ and $S$ is the
rescaled distance in the tangential direction
with ${\bf \hat{S}} \cdot \mbox{\boldmath{$\nabla$}}_{\bf X} =
\ell \partial_{S}$.
${\cal K}_{W} = \mbox{\boldmath{$\nabla$}}_{\bf X} \cdot
\hat{\bf w}$ is the curvature of the $W = constant$ contours and
${\cal K}_{S} = \mbox{\boldmath{$\nabla$}}_{\bf X} \cdot \hat{\bf s}$  is
the curvature of the $S = constant$ contours.
The time derivative is
$$
	\partial_{t} =
	\frac{ \partial \theta }{ \partial t }
	\partial_{ \theta } + \epsilon^{2} \partial_{T}
	=
	\epsilon \frac{\partial \Theta}{\partial T} \partial_{ \theta } +
	\epsilon^{2} \partial_{T}.
$$
We expand $\phi$ and $u$ in orders of $\epsilon$ as
\begin{equation}
	\phi = \phi_{0} + \epsilon \phi_{1} + {\cal O}(\epsilon^{2} ), \; \;
	u = u_{0} + \epsilon u_{1} + {\cal O}(\epsilon^{2} )
\end{equation}
We can now write the dynamical equations order by order in $\epsilon$.

\subsection{Zeroth Order Equations}

The zeroth order equations are
\begin{eqnarray}
	0 & = & - \mu_{B}(\phi_{0})
		+ k^{2} \partial_{\theta}^{2} \phi_{0} + u_{0},
	\nonumber \\
	0 & = &
		D k^{2} \partial_{\theta}^{2} u_{0}  - r_{0} \phi_{0}.
\end{eqnarray}
Therefore we find that $\phi_{0}$ and $u_{0}$ correspond to static
one-dimensional solutions of the model equations.  Since the static
solutions exists for a band of wavenumber $k$,
$\phi_{0}$ and $u_{0}$ depend on the slow variables through the local
wavenumber $k$.

\subsection{First Order Equations}

The first order equations are
\begin{eqnarray}
	0 & = &
		{\cal L}_{0}
		\left[ \begin{array}{c} \phi_{1} \\ w_{1} \end{array} \right]
		+ {\cal L}_{1}
		\left[ \begin{array}{c} \phi_{0} \\ w_{0} \end{array} \right]
\end{eqnarray}
where
\begin{equation}
	{\cal L}_{0}
	=
	\left[ \begin{array}{cc}
		-\frac{\partial \mu}{\partial \phi_{0}}
		+ k^{2} \partial^{2}_{\theta}
		&
		1 \\
		-r_{0} 	&
		D k^{2} \partial^{2}_{\theta}
		\end{array} \right]
\end{equation}
and
\begin{equation}
	{\cal L}_{1} =
		\left[ \begin{array}{cc}
		-\frac{ \partial \Theta }{ \partial T } \partial_{ \theta }
		+ \left( 2 k^{2} \partial_{W} + k \frac{\partial k}{\partial W}
		+ k {\cal K}_{W} \right) \partial_{ \theta }
		& 0 \\
		-\ell \frac{\partial \Theta}{\partial T} \partial_{\theta}
		&
		-\frac{\partial \Theta }{\partial T } \partial_{\theta}
		+ D \left( 2 k^{2} \partial_{W} + k
		\frac{\partial k}{\partial W}
		+ k {\cal K}_{W} \right) \partial_{ \theta }
		\end{array}
		\right].
\end{equation}
We apply the solvability condition
$\langle \, \nu_{0} \, {\cal L}_{1} \psi_{0} \rangle = 0$ where $\psi
= ( \phi, w )$ and $\nu_{0}$ are the null eigenmodes of ${\cal L}_{0}$,
that is $ \nu_{0} {\cal L}_{0} = {\cal L}^{\dag} \nu_{0} = 0$.  For
${\cal L}_{0}$ the null eigenmode is
$$
	\nu_{0} = \left( \partial_{\theta} \phi_{0}, -\frac{1}{r_{0} }
	\partial_{ \theta } w_{0}
	\right).
$$
The solvability condition gives
\begin{eqnarray}
	0 	& = &
	\left(	-\frac{\partial \Theta }{\partial T } + k^{2} \partial_{W}
	+ k \frac{\partial k}{\partial W}
	+ k {\cal K}_{W} \right) \;
		\frac{1}{ 2 \pi }
	\int^{2 \pi}_{0} d \theta \;
	\left( \partial_{ \theta } \phi_{0}  \right)^{2}
	+ \frac{\ell}{r_{0}}
	\frac{\partial \Theta }{\partial T } \frac{1}{  2 \pi }
	\int^{2 \pi}_{0} d \theta \;
	( \partial_{\theta} w_{0} ) \,
	( \partial_{\theta}  \phi_{0} )
	\nonumber \\
	& & +  \left(\frac{\partial \Theta }{\partial T }
	- D k^{2} \frac{\partial}{\partial w}
	- D k \frac{\partial k}{\partial w}
	- D k {\cal K}_{W} \right)
	\frac{1}{ 2 \pi }
	\int^{2 \pi}_{0} d \theta \;
	\left( \partial_{\theta} u_{0}  \right)^{2}.
\end{eqnarray}
We can introduce the coefficients
\begin{eqnarray}
	\frac{1}{ 2 \pi } \int^{2 \pi}_{0} \phi_{0}^{2} & = & A_{\phi}(k),
		\nonumber \\
	\frac{1}{ 2 \pi } \int^{2 \pi}_{0}
	\left( \partial_{ \theta } \phi_{0} \right)^{2} & = & A_{\phi}(k)
		q_{\phi,k},
		\nonumber \\
	\frac{1}{ 2 \pi } \int^{2 \pi}_{0} w_{0}^{2} & = & A_{u}(k),
		\nonumber \\
	\frac{1}{ 2 \pi } \int^{2 \pi}_{0}
	\left( \partial_{ \theta } w_{0} \right)^{2} & = & A_{u}(k) q_{u,k}.
		\nonumber
\end{eqnarray}
These coefficients depends only on the static one dimensional solutions
$\phi_{0}$ and $u_{0}$.  Therefore the coefficients depend on
the local wavenumber $k$.
We also have
\begin{eqnarray}
	\frac{1}{ 2 \pi } \int^{ 2 \pi }_{0} d \theta \;
	( \partial_{\theta} w_{0} ) \;
	( \partial_{ \theta } \phi_{0} )
	& = &
	-\frac{1}{ 2 \pi } \int^{ 2 \pi }_{0} d \theta \;
	\phi_{0} \; \partial^{2}_{\theta} w_{0},
	\nonumber \\
	& = &
	-\frac{r_{0}}{ D k^{2} }
	\frac{1}{ 2 \pi } \int^{ 2 \pi }_{0} d \theta \; \phi_{0}^{2},
		\\
	& = & -\frac{r_{0}}{ D k^{2} } A_{\phi}(k)
		\nonumber
\end{eqnarray}
Note that this implies that $A_{u}(k) \sim r_{0}^{2}/(D^{2} k^{4})$, confirming
our intuition that the temperature field mediates an effective long range
interaction.

\subsection{Phase Diffusion Equations and Liapunov Functional}

The phase diffusion equation is now
\begin{eqnarray}
	& & \left( A^{2} q_{\phi,k} + \frac{\ell}{ D k^{2} }  A_{\phi}(k)
	- \frac{1}{ r_{0} } A_{u}(k) q_{u,k} \right) \partial_{T} \Theta
		\nonumber \\
	&  = & \left[ k^{2} \partial_{W}
	+ k \frac{ \partial k}{\partial w} + k {\cal K}_{W} \right]
	\left( A_{\phi}(k) q_{\phi,k} - \frac{D}{r_{0}} A_{u}(k) q_{u,k} \right)
\end{eqnarray}
which we can write as
\begin{equation}
	\tau_{k} \partial_{T} \Theta
	=
	-\mbox{\boldmath{$\nabla$}}_{X} \cdot {\bf k} \, G(k).
\end{equation}
where
\begin{equation}
	G(k) = \frac{D}{r_{0}} A_{u}(k) q_{u,k}
		- A_{\phi}(k) q_{\phi,k}.
\end{equation}
and
\begin{equation}
	\tau_{k}
	=
	\left( A_{\phi}(k) q_{\phi,k} + \frac{\ell}{ D k^{2} } A_{\phi}(k)
	- \frac{1}{ r_{0} } A_{u}(k) q_{u,k} \right)
\end{equation}
In terms of the original coordinates the phase diffusion equation becomes
\begin{eqnarray}
	\tau_{k} \partial_{t} \theta
	& = &
	-\mbox{\boldmath{$\nabla$}} \cdot {\bf k} \, G(k), \nonumber
	\\
	& = &
	-G(k) \mbox{\boldmath{$\nabla$}} \cdot {\bf k}
	- k \frac{ d G(k) }{ d k } \frac{\partial k }{\partial w },
		\nonumber \\
	& = &
	-\frac{ d k G(k)}{d k}  \partial_{w}^{2} \theta
	- G(k) \partial_{s}^{2} \theta,
\end{eqnarray}
where we have used ${\bf k} = \mbox{\boldmath{$\nabla$}} \theta$.
Therefore, assuming $\tau_{k} > 0$,
the system is stable against variations in the normal
direction if $d (k G(k))/dk <0$ (the Eckhaus instability) and
stable against variations along the tangential direction if
$G(k) < 0$.

Now let us show that the selected wavelength is exactly the one marginally
stable against the zig-zag instability, i.e., $G(k) = 0$.  In the absence
of defects and boundary effects,
we can define a Liapunov function via \cite{CROSS}
\begin{equation}
	F\{ k( {\bf r} ) \}
	= -\frac{1}{2} \int d {\bf r}
	\int^{k( {\bf r})^{2} }_{0} d k'^{2} \; G(k').
\end{equation}
We can show $F$ is a Liapunov functional via
\begin{eqnarray}
	\partial_{t} F & = &
		-\frac{1}{2} \int d {\bf r} \; \partial_{t} k^{2} \;  G(k)
		=
		-\int d {\bf r} \; \tau_{k} \left( \partial_{t} \theta
		\right)^{2}.
\end{eqnarray}
It is easy to see that this function is minimised when $G(k) = 0$.

\subsection{Sharp Interface Limit}

We now calculate the coefficients in the
sharp interface limit.  Using the static solutions
equations (\ref{eq:PHISTRONG}) and (\ref{eq:USTRONG}) give
\begin{eqnarray}
	A_{\phi}(k) & = & 1 + {\cal O}(\epsilon) \nonumber \\
	A_{\phi}(k) q_{\phi,k} & = &
	\frac{1}{ \pi k} \bar{\sigma} + {\cal O}(\epsilon)
		 \\
	A_{u}(k) & = &
	\left( \frac{r_{0}}{D} \right)^{2} \frac{ \pi^{4}}{ 120 k^{4} }
	- \frac{17}{ 5040 } \left( \frac{r_{0}}{D} \right)^{3}
	\frac{\chi \pi^{6}}{ k^{6} }
	+ {\cal O}(\epsilon^{4})
	\nonumber  \\
	A_{u}(k) q_{u,k}
	& = &
	\left( \frac{r_{0}}{D} \right)^{2}
	\frac{ \pi^{2} }{ 12 k^{4} }
	-
	\frac{1}{30} \left( \frac{r_{0}}{D} \right)^{3}
	\frac{\chi \pi^{4}}{ k^{6} } + {\cal O}( \epsilon^{4} ).
		\nonumber
\end{eqnarray}
Note that $r_{0}$ is ${\cal O}(\epsilon^{3})$.
This gives
\begin{eqnarray}
	G(k) = \frac{r_{0}}{D} \frac{\pi^{2} }{12  k^{4} }
	-
	\frac{1}{60} \left( \frac{r_{0}}{D} \right)^{2}
	\frac{\chi \pi^{4}}{ k^{6} }
			 - \frac{\sigma}{\pi k}
			+ {\cal O}(\epsilon).
\end{eqnarray}
The leading order result for $\tau_{k}$ is
\begin{equation}
	\tau_{k}
	 =
	 \left( \frac{D - 1}{D} \right) \frac{\sigma}{\pi k}
	 +
	\frac{\ell}{ D k^{2} } - G(k) + {\cal O}(\epsilon).
\end{equation}
For $\tau_{k} > 0$, the stability condition against the zig-zag
instability is $G(k) < 0$ or
$\lambda < \lambda^{*}$ with
\begin{equation}
	\lambda^{*} = \left( \frac{ 96 D \bar{\sigma} }{ r_{0} } \right)^{1/3}
		+ \frac{ 8 \chi \bar{\sigma} }{ 5 }
		+ {\cal O}(\epsilon).
\end{equation}
For $\tau_{k} > 0$, the stability condition against the Eckhaus
instability is $ d (k G(k))/dk < 0$ or
\begin{equation}
	\frac{ d k G(k) }{ d k }
	= - \frac{ \pi^{2} r_{0} }{ 4 D k^{4} }
	+ \frac{ 2\pi^{6} \chi r_{0}^{2} }{ 3 D^{2} k^{6} }
	+ {\cal O}(\epsilon) < 0.
\end{equation}
Therefore if the lamellae are stable against the zig-zag instability,
they are also stable to the Eckhaus instability.

\subsection{Weak segregation limit}

In the weak segregation limit, we expand in $\epsilon^{2} = B_{0} - B$.
Using the static solution given by equations (\ref{eq:PHIWEAK}) and
(\ref{eq:UWEAK}), the coefficients become
\begin{eqnarray}
	A_{\phi}(k) & = &
		\frac{\epsilon^{2}  A_{k}^{2} }{2}
	\left( 1 + \epsilon^{4} \left( \frac{9}{128} \right)^{2}
	A_{k}^{4} \right),
		\nonumber \\
	A_{\phi}(k) q_{\phi,k} & = &
		\frac{ \epsilon^{2} A_{k}^{2} }{2}
		\left( 1 + \epsilon^{4} \left( \frac{27}{128} \right)^{2}
		A_{k}^{4} \right),
		\nonumber \\
	A_{u}(k) & = &
	\frac{\epsilon^{2}  A_{k}^{2} }{2}
	\frac{r_{0}^{2} }{D^{2} k^{4}}
	\left( 1 + \epsilon^{4} \left( \frac{1}{128} \right)^{2}
	A_{k}^{4} \right),
		\\
	A_{u}(k) q_{u,k} & = &
		\frac{\epsilon^{2}  A_{k}^{2} }{2}
		\frac{ r_{0}^{2} }{D^{2} k^{4} }
		\left( 1 + \epsilon^{4} \left( \frac{3}{128} \right)^{2}
		A_{k}^{4} \right).
		\nonumber
\end{eqnarray}
This gives $G(k)$ as
\begin{equation}
	G(k) = \frac{\epsilon^{2}  A_{k}^{2} }{2}
	\left( 1 - \frac{r_{0}}{D k^{4} } +
	\epsilon^{4} \left( \frac{3}{128} \right)^{2} A_{k}^{4}
	\left( 81 - \frac{r_{0}}{D k^{4} } \right) \right).
\end{equation}
The leading order expression for $\tau_{k}$ is
\begin{equation}
	\tau_{k}
	=
	\epsilon^{2}  A_{k}^{2} \frac{ r_{0} }{D } \frac{D-1}{2 D k^{4} }
	+ \epsilon^{2} A_{k}^{2} \frac{ \ell }{2 D k^{2} }
	- G(k)  + {\cal O}(\epsilon^{6}).
\end{equation}

\section{Numerical details}

In order to numerically update the laser induced melting model we
use a Euler discretization.  Due to the large mesh size and time
step ($\delta x = 1.25$ and $\delta t = 0.2$) the update does
not reproduce the partial
differential equations in the sense of obtaining the exact position
of interfaces.  However, the numerical method does reproduce
the large scale behaviour such as the average pattern length.  This is
checked qualitatively by varying the mesh sizes and step sizes.

The mesh can produce large anisotropy
effects.  Oono and Puri reduced the
anisotropy by including the next nearest neighbours in the
discretized Laplacian
\cite{OONOCDS}.  They used for their Laplacian (in 2-$d$)
\begin{eqnarray}
	(\delta x)^{2} [ \nabla^{2} u ]_{i,j}
		& = &
	\frac{1}{ 1 + 2 \Delta }
	\Bigr[ u_{i+1,j} + u_{i,j+1} + u_{i-1,j} + u_{i,j-1}
	\nonumber \\
	& &
	+ \Delta \left( u_{i+1,j+1} + u_{i+1,j-1} + u_{i-1,j-1} +
			u_{i+1,j-1} \right)
	- 4 ( 1 + \Delta ) u_{i,j}
	\bigr],
\end{eqnarray}
with $\Delta = 1/2$ so that the next nearest neighbours 1/2 weight
relative to the nearest neighbours.

Here we consider different values of $\Delta$.
In fourier representation
the discretized Laplacian is
\begin{eqnarray}
	(\delta x)^{2} \, \Gamma({\bf k}) &  = &
		\frac{2}{ 1 + 2 \Delta }
		\Bigr[ 1 - \cos( k_{x} \delta x )  + 1 - \cos( k_{y} \delta x )
			\nonumber \\
	& &
		+ \Delta \, \left( 1 - \cos( (k_{x}+k_{y}) \delta x )
		+ 1 - \cos( ( k_{x} - k_{y} ) \delta x ) \right) \Bigr].
\end{eqnarray}
We want to make $\Gamma_{k}$ as isotropic as possible. As a measure
of the anisotropy we use
the ratio of $\Gamma( {\bf k})$ at different
angles $\theta$ where ${\bf k} = ( k \cos( \theta ), k \sin( \theta ) )$.

Figure \ref{fig:RATIO} shows a plot of the ratio
$\Gamma(k/\sqrt{2},k/\sqrt{2})/\Gamma(k,0)$ for $\Delta
= 0, 0.3, 0.35$ and $1/2$. For all values of $\Delta$,
this ratio approaches unity in the limit
$k \delta x \rightarrow 0$.
For small $k \delta x$, the ratio is larger than one and
becomes unity at a value of $k \delta x = (k \delta x)^{*}$ which
increases with increasing $\Delta$.  Note that $(k \delta x)^{*} =
0$ for $\Delta = 0$.  From this plot it is clear that the choice of
$\Delta = 1/2$ does reduce the anisotropy from $\Delta = 0$.  However,
$\Delta = 1$ makes $\Gamma$ on the diagonals too large relative to
the axis.  For our simulations we choose $\Delta = 0.35$
which seem to give a nice compromise between making the value of
$\Gamma$ on the diagonals too large for small $k \delta x$ and
too small for larger $k \delta x$.

\begin{figure}
\caption{
The $\phi$ field for $L_{x} = L_{y} = 640$, $dx = 1.25$, $L=2.0$, $D=0.5$,
and $\Delta j = 0$.  A $256 \times 256$ portion of the $512 \times 512$
lattice is shown.  From top to bottom, the left panels
shows $r_{0}/D = 0.24$ (close to onset)
with $t$ = 1600, 12800 and 102400.  The rolls continue to straighten at
later times.  The right shows $r_{0}/D = 0.001875$ (far from the onset)
for the same times.  The patterns are essentially frozen after this time.
\label{fig:PICTURE}
}
\end{figure}

\begin{figure}

\caption{
Graphical summary of the stability analysis.  The homogeneous state
is linearly unstable within the shaded area.  The solid line is $k^{*}
= (r_{0}/D)^{1/4}$, i.e., the lowest order result for the selected wavenumber
obtained near onset.  The dotted line is the lowest order wavenumber
far from onset, $k^{*} = 2 \pi ( 96 D \bar{\sigma} / r_{0} )^{-1/3}$.
The dashed line is the far from onset result including next order
corrections.  The squares are the length-scale obtained from simulations
with $\ell = 2$ and $D = 0.5$.
\label{fig:LENGTHS}
}
\end{figure}

\begin{figure}
\caption{
\label{fig:PROFILE}
The values of the $\phi$ and $u$ field along a horizontal cut for two
values of $r_{0}/D$.
The dashed line with the
solid circles is the $\phi$ field while the solid line with open boxes
is the $u$ field.
(a)
$r_{0}/D = 0.24$ (near onset).  (b) $r_{0}/D = 0.001875$ (far from onset).
}
\end{figure}

\begin{figure}
\caption{
The circular averaged scattering intensities $S_{k}(t)$ for $t$ =
1600, 12800 and 102400.  (a) $r_{0}/D = 0.24$ (close to
onset).  The sharp peak in $S_{k}(t)$ is fixed at $k \approx 0.7$.
(b) $r_{0}/D = 0.001875$.
The broad peak shifts toward lower $k$ with time.
\label{fig:SCATTERING}
}
\end{figure}

\begin{figure}
\caption{
Test of the scaling form $k_{1}(t) = k_{1}(\infty) g( t r_{0}/D )$.
Plot is of $k_{1}(t)/k_{1}(\infty)$ vs. $r_{0}/D$ for seven values
of $r_{0}/D$ from $0.00046875$ to $0.24$.  The line is of the form $k
\sim t^{1/3}$.  The inset shows the unscaled plot of $k_{1}(t)$ vs. $t$.
\label{fig:SCALING}
}
\end{figure}

\begin{figure}
\caption{
The director field for $r_{0}/D = 0.24$ corresponding to the same
times as in Figure 1 ($t=1600, 12800$ and $102400$.).  The entire
$512 \times 512$ lattice is shown.  The lower left quarter box
corresponds to $256 \times 256$ region shown in Figure 1.  White
denotes regions with $\cos(2 \theta) > 0$ ($-\pi/2 < 2 \theta < \pi/2$)
and black denotes  $\cos(2 \theta) < 0$.
\label{fig:OPICTURE}
}
\end{figure}

\begin{figure}
\caption{
The orientational correlation function $C_{\theta}$.  (a) $C_{\theta}$
near onset ($r_{0}/D = 0.24$) with times $t=800$ $(-)$,
1600 ($\circ$), 3200 ($...$), 4800 ($\bullet$), 6400
(---) and 12800 ($*$).  The inset is
the region near $x = 0$.  (b) $C_{\theta}$ further from onset
($r_{0}/D = 0.001875$).  Times $t$ = 800, 3200, 12800 and 51200 are shown.
\label{fig:ORIENTATION}
}
\end{figure}

\begin{figure}
\caption{
The orientational correlation length $L_{\theta}(t)$ for
$r_{0}/D = 0.06$, $0.12$, $0.18$ and $0.24$ on log-log scale.
A line of $L \sim t^{1/4}$ is also drawn. The data for $r_{0}/D =
0.24$ is strongly influenced by finite size effects.
\label{fig:LTHETAONSET}
}
\end{figure}

\begin{figure}
\caption{
The orientational correlation length $L_{\theta}  k_{1}$ for
8 values of $r_{0}/D$ ranging from
$r_{0}/D = 0.00046875$ to $r_{0}/D= 0.24$ on log-log scale.
This product is expected to be approaching its asymptotic values except for two
points closest to onset (open squares).
\label{fig:LTHETAK}
}
\end{figure}

\begin{figure}
\caption{
The value of $\ell$ at which the transition from stripe phase
to oscillating homogeneous phase occurs as a function of $r_{0}$.
The error bars indicate the bracketed region as explained in
the text.  The shaded area is where the
linear stability analysis predicts oscillatory behavior.
The solid line is the value of $\ell$ at which the simultaneous
Eckhaus/zig-zag instability occurs obtained from the weak
segregation limit and the dashed line is that value
obtained from the strong segregation limit.  A line proportional
to $r_{0}^{1/2}$ is drawn as a guide to the eye.
\label{fig:CHTRANSITION}
}
\end{figure}

\begin{figure}
\caption{
The ratio of the sphericalized Laplacian in Fourier space along the diagonals
divided by its value along the axis.  $\Delta = 0$, $\Delta = 0.3$, $\Delta =
0.35$ and $\Delta = 1/2$ are shown.
\label{fig:RATIO}
}
\end{figure}

\end{document}